\documentclass[aps,prb,reprint,showpacs,groupedaddress,floatfix]{revtex4-1}
\usepackage{graphicx}
\usepackage{color}
\usepackage{amsmath,amssymb}
\usepackage[caption=false]{subfig}
\usepackage{hyperref}

\newcommand{\dd}{d} % APS uses italic derivative
\newcommand{\du}[1]{\dd#1}
\providecommand*{\deriv}[3][]{
    \frac{\dd^{#1}#2}{\dd#3^{#1}}}
\providecommand*{\pderiv}[3][]{
    \frac{\partial^{#1}#2}{\partial#3^{#1}}}

\newcommand{\unit}[1]{\mathrm{#1}}
\newcommand{\bvec}[1]{\boldsymbol{\mathrm{#1}}}
\newcommand{\buvec}[1]{\hat{\bvec{#1}}} % bold unit bvecor
\newcommand{\eu}{\ensuremath{e}} % APS uses italic e
\newcommand{\iu}{\ensuremath{i}}
\renewcommand\Re{\operatorname{\mathfrak{Re}}}
\renewcommand\Im{\operatorname{\mathfrak{Im}}}

\newcommand{\gTensorXY}{\widehat{\mathcal{G}}_{xy}}
\newcommand{\oH}{\omega_H}
\newcommand{\oM}{\omega_M}
\newcommand{\oHExplicit}{\gamma H_i}
\newcommand{\oMExplicit}{4\pi\gamma M_s}
\newcommand{\ooM}{\frac{\omega}{\oM}}
\newcommand{\oHoM}{\frac{\oH}{\oM}}

\newcommand{\dk}{\delta k}
\newcommand{\dEff}{d_\text{eff}}

\newcommand{\lEx}{\lambda_\text{ex}}

\newcommand{\lExExplicitInline}{\sqrt{8\pi\gamma^2A/\oM^2}}

\newcommand{\gMix}{g_\perp}
\newcommand{\gMixTilde}{\tilde{g}_\perp}
\newcommand{\pPar}{\rho}
\newcommand{\pParTilde}{\tilde{\rho}}

\newcommand{\dynMQ}{\bvec{m}_{Q}}
\newcommand{\Mtotal}{\bvec{M}(\bvec{r},t)}
\newcommand{\Heff}{\bvec{H}_\text{eff}}

\newcommand{\muS}{\mu^{(s)}}

\newcommand{\tsf}{\tau_\text{sf}}
\newcommand{\lsf}{l_\text{sf}}
\newcommand{\lsfj}{l_{\text{sf},j}}
\newcommand{\lsfAC}{l_{\text{sf},AC}}

\newcommand{\mcomment}[1]{} % multiline commenting
\begin{document}

\title{Spin Pumping, Dissipation, and Direct and Alternating Inverse Spin Hall
Effects in Magnetic Insulator-Normal Metal Bilayers}

\author{Andr\'{e} Kapelrud}
\author{Arne Brataas}
\affiliation{Department of Physics, Norwegian University of Science and Technology,
NO-7491 Trondheim, Norway}


\begin{abstract}
We theoretically consider the spin-wave mode- and wavelength-dependent enhancement of
    the Gilbert damping in magnetic insulator—normal metal bilayers due to spin
    pumping as well as the enhancement's relation to direct and alternating inverse
    spin Hall voltages in the normal metal. In the long-wavelength limit, including
    long-range dipole interactions, the ratio of the enhancement for transverse
    volume modes to that of the macrospin mode is equal to two. With an out-of-plane
    magnetization, this ratio decreases with both an increasing surface anisotropic
    energy and mode number. If the surface anisotropy induces a surface state, the
    enhancement can be an order of magnitude larger than for to the macrospin.  With
    an in-plane magnetization, the induced dissipation enhancement can be understood
    by mapping the anisotropy parameter to the out-of-plane case with anisotropy.
    For shorter wavelengths, we compute the enhancement numerically and find good
    agreement with the analytical results in the applicable limits. We also compute
    the induced direct- and alternating-current inverse spin Hall voltages and relate
    these to the magnetic energy stored in the ferromagnet. Because the magnitude of
    the direct spin Hall voltage is a measure of spin dissipation, it is directly
    proportional to the enhancement of Gilbert damping. The alternating spin Hall
    voltage exhibits a similar in-plane wave-number dependence, and we demonstrate
    that it is greatest for surface-localized modes.
\end{abstract}

\maketitle
%\tableofcontents
%--------------------------------------------------------------------
\section{Introduction\label{sec:introduction}}
In magnonics, one goal is to utilize spin-based systems for interconnects and logic
circuits \cite{Serga2010.43.264002}. In previous decades, the focus was to gain
control over these systems by exploiting long-range dipole interactions in
combination with geometrical shaping. However, the complex nature of the nonlinear
magnetization dynamics persistently represents a challenge in using geometrical
shaping alone to realize a variety of desired properties\cite{Serga2010.43.264002}.

In magnonic systems, a unique class of materials consists of magnetic insulators.
Magnetic insulators are electrically insulating, but localized magnetic moments
couple to form a long-range order. The prime example is \textit{Yttrium Iron Garnet}
(YIG). YIG is a complex crystal\cite{Cherepanov199381} in the Garnet family, where
the $\unit{Fe}^{2+}$ and $\unit{Fe}^{3+}$ ions at different sites in the unit cell
contribute to an overall ferrimagnetic ordering. What differentiates YIG from other
ferromagnetic (ferrimagnetic) systems is its extremely low intrinsic damping. The
Gilbert damping parameter measured in YIG crystals is typically two orders of
magnitude smaller than that measured in conventional metallic ferromagnets (Fe, Co,
Ni, and alloys thereof).

The recent discovery that the spin waves in magnetic insulators strongly couple to
spin currents in adjacent normal metals has re-invigorated the field of
magnonics\cite{Nature.464.262,APL.97.252504,PhysRevLett.106.216601,
PhysRevLett.107.066604,rezende:012402,Rezende:apl2011,PhysRevLett.109.026602,
burrowes:092403,Hillebrands.gMix.2013,PhysRevLett.111.217204}. Although there are no
mobile charge carriers in magnetic insulators, spin currents flow via spin waves and
can be transferred to itinerant spin currents in normal metals via spin transfer and
spin pumping\cite{RevModPhys.77.1375,Brataas:NatMat2012}. These interfacial effects
open new doors with respect to local excitation and detection of spin waves in
magnonic structures. Another key element is that we can transfer knowledge from
conventional spintronics to magnonics, opening possibilities for novel physics and
technologies. Traditionally, spin-wave excitation schemes have focused on the
phenomenon of resonance or the use of \O rsted fields from microstrip antennas.

A cornerstone for utilizing these systems is to establish a good understanding of how
the itinerant electrons in normal metals couple across interfaces with spin-wave
dynamics in magnetic insulators. Good models for adressing uniform (macrospin)
magnetization that agrees well with experiments have been previously
developed\cite{Brataas:NatMat2012,PhysRevLett.88.117601,RevModPhys.77.1375}. We
recently demonstrated that for long-wavelength magnons the enhanced Gilbert damping
for the transverse volume modes is twice that of the uniform mode, and for surface
modes, the enhancement can be more than ten times stronger. These results are
consistent with the theory of current-induced excitations of the magnetization
dynamics\cite{PhysRevLett.108.217204} because spin pumping and spin transfer are
related by Onsager reciprocity relations\cite{maekawa2012spin}. Moreover, mode- and
wave-vector-dependent spin pumping and spin Hall voltages have been clearly observed
experimentally\cite{APL.97.252504}.

In this paper, we extend our previous findings\cite{PhysRevLett.111.097602} in the
following four aspects. i) We compute the influence of the spin backflow on the
enhanced spin dissipation. ii) We also compute the induced direct and alternating
inverse spin Hall voltages. We then relate these voltages to the enhanced Gilbert
damping and the relevant energies for the magnetization dynamics. The induced
voltages give additional information about the spin-pumping process, which can also
be directly measured. iii) We also provide additional information on the effects of
interfacial pinning of different types in various field geometries. iv) Finally, we
explain in more detail how the numerical analysis is conducted for a greater number
of in-plane wave numbers.

It was discovered\cite{PhysRevLett.29.1327, PhysRevB.19.4382, PhysRevLett.37.612,
mizukami2001study, PhysRevLett.87.217204} and later quantitatively
explained\cite{PhysRevLett.88.117601, PhysRevB.66.060404, PhysRevB.66.224403,
RevModPhys.77.1375} that if a dynamic ferromagnetic material is put in contact with a
normal metal, the magnetization dynamics will exert a torque on the spins of electrons
in the immediate vicinity of the magnet. This effect is known as spin pumping
(SP)\cite{PhysRevLett.88.117601,PhysRevB.66.224403,RevModPhys.77.1375}. As the
electrons are carried away from the ferromagnet-normal metal interface, the electrons
spin with respect to each other, causing an overall loss of angular momentum. The
inverse effect, in which a spin-polarized current can affect the magnetization of a
ferromagnet, is called spin-transfer torque (STT)\cite{Berger1996, Slonczewski1996,
brataas2012current}.

The discovery that a precessing magnetization in magnetic
insulators\cite{Nature.464.262}, such as YIG, also pumps spins into an adjacent metal
layer was made possible by the fact that the mixing conductance in YIG-normal metal
systems is of such a size that the extra dissipation of the magnetization due to the
spin pumping is of the same order of magnitude as the intrinsic Gilbert damping. A
consequence of this effect is that the dissipation of the magnetization dynamics is
enhanced relative to that of a system in which the normal metal contact is removed.

This paper is organized in the following manner. Section \ref{sec:theory} presents
the equation of motion for the magnetization dynamics and the currents in the normal
metal and the appropriate boundary conditions, both for general nonlinear excitations
and in the fully linear response regime. In Section \ref{sec:sptheory}, we derive
approximate solutions to the linearized problem, demonstrating how the magnetization
dissipation is enhanced by the presence of an adjacent metal layer. Section
\ref{sec:numerics} presents our numerical method and results. Finally, we summarize
our findings in Section \ref{sec:conclusion}.

\section{Equations of Motion\label{sec:theory}}
The equation of motion for the magnetization is given by the Landau-Lifshitz-Gilbert
equation\cite{Gilbert1955} (presented here in CGS units)
\begin{equation}
    \pderiv{\bvec M}{t}=
        -\gamma\bvec M\times\Heff
        +\frac{\alpha}{M_s}\bvec M\times\pderiv{\bvec M}{t},
    \label{eq:LLG}
\end{equation}
where $\gamma=|g\mu_B/\hbar|$ is the magnitude of the gyromagnetic ratio; $g\approx
2$ is the Land\'{e} g-factor for the localized electrons in the ferromagnetic
insulator (FI); and $\alpha$ is the dimensionless Gilbert damping parameter. In
equilibrium, the magnitude of the magnetization is assumed to be close to the
saturation magnetization $M_s$. The magnetization is directed along the $z$-axis in
equilibrium. Out of equilibrium, we assume that we have a small transverse dynamic
magnetization component, such that
\begin{equation}
    \bvec M=\Mtotal=\bvec M_s+\bvec m(\bvec r,t)=M_s\buvec z+\bvec m(\bvec r,t),
    \label{eq:magnetization}
\end{equation}
where $|\bvec m|\ll M_s$ and $\bvec m\cdot\buvec z=0$.
%--------------------------------------------------------------------
\begin{figure}[bthp]
\centering
    \subfloat[][]{
        \includegraphics{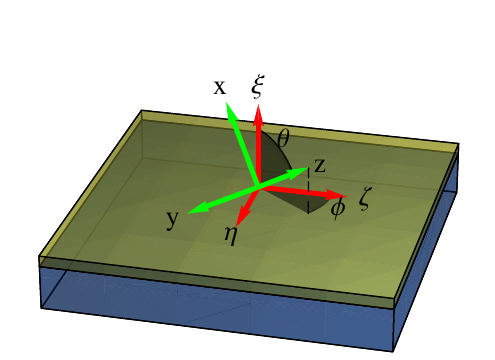} \label{fig:angles}
    }
    \subfloat[][]{
        \includegraphics{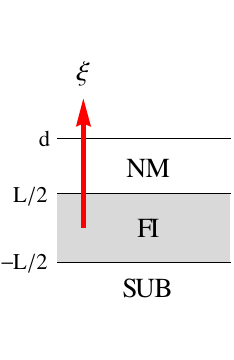} \label{fig:cross_section}
    }
    \caption{\label{fig:geometry}a) The coordinate system. $\buvec\xi$ is the film
    normal and $\buvec\zeta$ is the spin-wave propagation direction. $\xi\eta\zeta$
    form a right-handed coordinate system. The $\buvec z$ axis is the direction of
    the magnetization in equilibrium, such that $xy$ is the magnetization-precession
    plane. b) The film stack is in the normal direction.}
\end{figure}
%--------------------------------------------------------------------
Furthermore, we assume that the dynamic magnetization can be described by a plane
wave traveling along the in-plane $\zeta$-axis. In the $(\xi,\eta,\zeta)$ coordinate
system (see Figure~\ref{fig:geometry}), we have
\begin{equation}
    \bvec m(\bvec r,t)=\bvec m(\xi,\zeta,t)=\bvec m_{Q}(\xi)\eu^{\iu(\omega
    t-Q\zeta)},
    \label{eq:magn_xizeta}
\end{equation}
where $\omega$ is the harmonic angular frequency, $Q$ is the in-plane wave
number, and $\bvec m_Q(\xi)=X_Q(\xi)\buvec x+Y_Q(\xi)\buvec y$, where $X_Q$ and $Y_Q$
are complex functions. Note that $\bvec m$ is independent of the $\eta$ coordinate
due to translational invariance.

$\Heff$ is the effective field, given as the functional derivative of the free energy
\cite{Gilbert1955,Gilbert2004}
\begin{align}
    \Heff(\bvec r,t)= &-\frac{\delta U[\Mtotal]}{\delta\Mtotal}
        %= \bvec H_{i}+\bvec H_{ex}+\bvec h_{d}\nonumber\\
    = \bvec H_{i}+\frac{2A}{M_s^2}\nabla^2\bvec M(\bvec r,t)+\nonumber\\
    &\quad+4\pi\int_{-\frac{L}{2}}^{\frac{L}{2}}\dd\xi'\,\gTensorXY(\xi-\xi')
    \bvec m(\xi',\zeta,t),
    \label{eq:Heff}
\end{align}
where $\bvec H_{i}$ is the internal field, which is composed of the applied external
field and the static demagnetization field. The direction of $\bvec H_i$ defines the
$z$-axis (see Figure~\ref{fig:geometry}). The second term of Eq.~\eqref{eq:Heff} is
the field, $\bvec H_{ex}$, induced by to the exchange interaction (assuming cubic symmetry),
where $A$ is the exchange stiffness parameter. The last term is the dynamic field,
$\bvec h_{d}(\bvec r,t)$, induced by dipole-dipole interactions, where $\gTensorXY$ is
the upper $2\times2$ part of the dipole--dipole tensorial Green's function
$\widehat{\mathcal{G}}_{\xi\eta\zeta}$ in the magnetostatic approximation\citep[see
Ref.~][]{Kalinikos1981} rotated to the $xyz$ coordinate system (see Appendix
\ref{app:coord-trans} for coordinate-transformation matrices).\cite{Kalinikos1986}

\begin{figure*}[hbtp]
\centering
    \subfloat[][]{
        \includegraphics{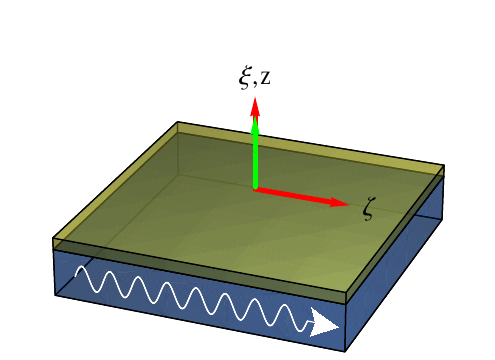} \label{fig:geomFVMSW}
    }
    \subfloat[][]{
        \includegraphics{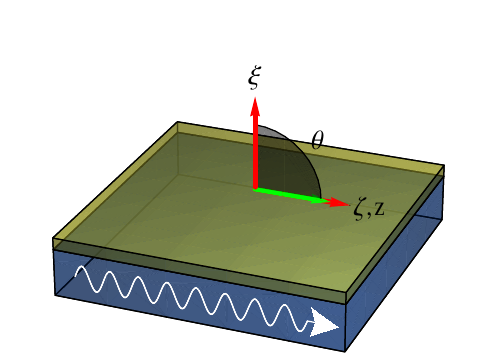} \label{fig:geomBVMSW}
    }
    \subfloat[][]{
        \includegraphics{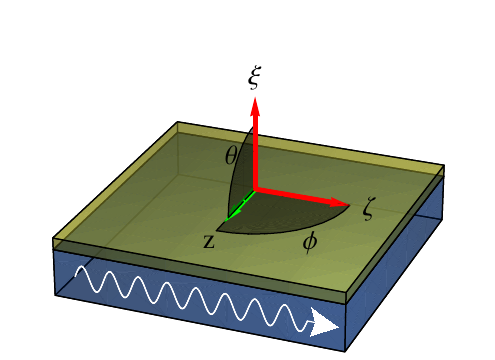} \label{fig:geomMSSW}
    }
    \caption{\label{fig:symmetries}Laboratory field configurations, i.e., directions
    of $\buvec z$ (green arrow) in relation to film normal $\buvec\xi$ and the
    spin-wave propagation direction $\buvec\zeta$, resulting in the different
    geometries: a) FVMSW geometry; b) BVMSW geometry; c) MSSW geometry.}
\end{figure*}

The effect of the dipolar interaction on the spin-wave spectrum depends on the
orientation of the internal field with respect to both the interface normals of the
thin film, $\buvec\xi$, and the in-plane spin-wave propagation direction,
$\buvec\zeta$. Traditionally, the three main configurations are the out-of-plane
configuration ($\theta=0$), in the \textit{forward volume magnetostatic wave} (FVMSW)
geometry (see Fig.~\ref{fig:geomFVMSW}); the in-plane and parallel-to-$\buvec\zeta$
configuration, in the \textit{backward volume magnetostatic wave} (BVMSW) geometry
(see Fig.~\ref{fig:geomBVMSW}); and the in-plane and perpendicular-to-$\buvec\zeta$
configuration, in the \textit{magnetostatic surface wave} (MSSW) geometry (see
Fig.~\ref{fig:geomMSSW}).\cite{PhysRev.118.1208, Damon1961308, Puszkarski.IEEE.1973,
wames:987, Kalinikos1986, Serga2010.43.264002} Here, the term ``forward volume
modes'' denotes modes that have positive group velocities for all values of $QL$,
whereas backward volume modes can have negative group velocities in the range of
$QL$, where both exchange and dipolar interactions are significant. Volume modes are
modes in which $\bvec m_Q(\xi)$ is distributed across the thickness of the entire
film, whereas the surface modes are localized more closely near an interface.
%--------------------------------------------------------------------
%--------------------------------------------------------------------

\subsection{Spin-Pumping Torque}
%--------------------------------------------------------------------
We consider a \textit{ferromagnetic insulator} (FI) in contact with a \textit{normal
metal} (NM) (see Figure~\ref{fig:geometry}). If the magnetization in the FI close to
the interface is precessing around the effective field, electron spins in the NM
reflected at the interface will start to precess due to the local exchange coupling
to the magnetization in the FI. The reflected electrons carry the angular momentum
away from the interface, where the spin information can get lost through dephasing of the
spins within a typical spin diffusion length $\lsf$. This loss of angular momentum
manifests itself as an increased local damping of the magnetization dynamics in the
FI. The magnetization dissipation due to the spin-pumping effect can be taken into
account by adding the local dissipation torque\cite{PhysRevLett.88.117601}
\begin{equation}
        \bvec\tau_\text{sp}=\frac{\gamma\hbar^{2}\gMix}{2e^{2}M_s^{2}}\delta(\xi-\frac{L}{2})
            \Mtotal\times\pderiv{\Mtotal}{t},
    \label{eq:tau_sp}
\end{equation}
to the right-hand side (rhs) of Eq.~\eqref{eq:LLG}. Here, $\gMix$ is the real part of
the spin-mixing conductance per area, and $e$ is the electron charge. We neglect the
contribution from the imaginary part of the mixing conductance, because this has been
shown to be significantly smaller than that of the real part, in addition to
affecting only the gyromagnetic ratio.\cite{PhysRevLett.88.117601} The spin-current
density pumped from the magnetization layer is thus given by \begin{equation}
    \bvec j^{(s)}_\text{sp}=
    -\frac{\hbar^2\gMix}{2e^2M_s^2}
    \left[\Mtotal\times\pderiv{\Mtotal}{t}\right]_{\xi=L/2},
    \label{eq:sp_current}
\end{equation}
in units of $\unit{erg}$. Next, we will see how the spin pumping affects the boundary
conditions.

\subsection{Spin-Pumping Boundary Conditions\label{sec:spin_pumping_bc}}
%-------------------------------------------------------------------------------
Following the procedure of \citet{RadoWeertman1959}, we integrate Eq.\eqref{eq:LLG}
with the linear expansion of Eq.~\eqref{eq:magnetization} over a small pill-box
volume straddling one of the interfaces of the FI. Upon letting the pill box
thickness tend to zero, only the surface torques of the equation survive. Accounting
for the direction of the outward normal of the lid on the different top and bottom
interfaces, we arrive at the exchange-pumping boundary condition
\begin{equation}
    \left[\frac{2A}{M_s^2}\bvec M\times\frac{\partial{\bvec M}}{\partial{\xi}}
    +\frac{\hbar^2}{2e^2 M_s^2}\gMix
     \bvec M\times\frac{\partial\bvec m}{\partial t}\right]_{\xi=\pm L/2}
    =0.\label{eq:exchange_pumping_BC}
\end{equation}

There is no spin current pumped at the interface to the insulating substrate; thus, a
similar derivation results in a boundary condition that gives an unpinned
magnetization,
\begin{equation}
    \left.\pderiv{\Mtotal}{\xi}\right|_{\xi=-L/2}=0.
    \label{eq:exchange_BC_lower}
\end{equation}

In the next section, we will generalize the boundary conditions of Eqs.\
\eqref{eq:exchange_pumping_BC} by also considering possible surface-anisotropy
energies.

\subsubsection*{Including surface anisotropy:}
%-------------------------------------------------------------------------------
In the presence of surface anisotropy at an interface with an \textit{easy-axis} (EA)
pointing along the direction $\buvec n$, the surface free energy is 
\begin{equation}
U_s[\Mtotal]=\int\du V\, K_s\left[
    1-\left(\frac{\Mtotal\cdot\buvec n}{M_s}\right)^{2}
    \right]\delta\left(\xi-\xi_i\right),
    \label{eq:UEA}
\end{equation}
where $K_s$ is the surface-anisotropy energy density at the interface, which is
assumed to be constant; $\buvec n$ is the direction of the anisotropy \textit{easy}
axis; and $\xi_i$ is the transverse coordinate of the interface. The contribution
from the EA surface-anisotropy energy to the effective field is determined by
\[
    \bvec H_s=
    -\frac{\delta U_s[\Mtotal]}{\delta\bvec \Mtotal}=
    \frac{2K_s}{M_s^2}\left(\bvec M\cdot\buvec n\right)
    \delta(\xi-\xi_{j})\buvec n.
\]
However, if we have an \textit{easy-plane} (EP) surface anisotropy with, $\buvec n$
being the direction of the \textit{hard} axis, the effective field is the same as
that for the EA case, except for a change of sign of $K_s$. We unify both cases by
defining $K_s>0$ to imply that we have an EA surface anisotropy with its easy axis
along $\buvec n$, whereas $K_s<0$ implies that we have an EP surface anisotropy with
its hard axis along $\buvec n$.

Following the approach from Section~\ref{sec:spin_pumping_bc}, the total boundary
condition, including exchange, pumping and surface anisotropy, becomes
\begin{multline}
\left[\pm\frac{2A}{M_s^2}\bvec M\times\frac{\partial{\bvec M}}{\partial{\xi}}
    -\frac{2K_s}{M_s^2}\left(\bvec M\cdot\buvec n\right)\left(\bvec M\times\buvec
    n\right)\right.+\\
    \left.+\frac{\hbar^2}{2e^2 M_s^2}\gMix\bvec M\times\frac{\partial\bvec M}{\partial t}
    \right]_{\xi=\pm L/2}=0,
    \label{eq:exPumpAni_BC}
\end{multline}
where the positive (negative) sign in front of the exchange term indicates that the
bulk FI is located below (above) the interface coordinate.

\subsection{Linearization\label{sec:linearization}}
%-------------------------------------------------------------------------------
We linearize the equation of motion using Eq.~\eqref{eq:magnetization} with respect
to the dynamic magnetization $\bvec m$. The linearized equation of motion for the
bulk magnetization Eq.~\eqref{eq:LLG} becomes\cite{Kalinikos1986}
\begin{multline}
    \left\{\iu\frac{\omega}{\oM}
    \begin{pmatrix}
        \alpha & -1 \\
        1 & \alpha
    \end{pmatrix}
    +\openone\left[
        \frac{\oH}{\oM}
            +8\pi\frac{\gamma^2A}{\oM^2}\left(Q^2-\deriv[2]{}{\xi}\right)
    \right]\right\}\cdot\\
    \cdot\dynMQ(\xi)
    =\int_{-\frac{L}{2}}^{\frac{L}{2}}\du{\xi'}\,\gTensorXY(\xi-\xi')\dynMQ(\xi'),
    \label{eq:motion}
\end{multline}
where $\oH\equiv\oHExplicit$, $\oM\equiv\oMExplicit$, and
$\openone=\big(\begin{smallmatrix}1&0\\0&1\end{smallmatrix}\big)$.

Next, we linearize the boundary conditions of Eq.~\eqref{eq:exPumpAni_BC}. We choose
the anisotropy axis to be perpendicular to the film plane, $\buvec n=\buvec \xi$,
which in the $xyz$ coordinate system is given by
$\buvec\xi_{xyz}=(\sin\theta,0,\cos\theta)$, where $\theta$ is the angle between the
$z$-axis and the film normal (see Fig.~\ref{fig:geometry}). The finite surface
anisotropy forces the magnetization to be either perpendicular or coplanar with the
film surface so that $\theta=0,\pi/2,\pi$. Linearizing to 1st order in the dynamic
magnetization, we arrive at the linearized boundary conditions for the top interface
\begin{subequations}
\label{eq:BClin}
\begin{gather}
    \left.\left(
        L\pderiv{}{\xi}+\iu\ooM\pPar+d\cos(2\theta)
    \right) m_{Q,x}(\xi)\right|_{\xi=\frac{L}{2}}=0,
    \label{eq:BClinX} \\
    \left.\left(
        L\pderiv{}{\xi}+\iu\ooM\pPar+d\cos^2(\theta)
    \right) m_{Q,y}(\xi)\right|_{\xi=\frac{L}{2}}=0,
    \label{eq:BClinY}
\end{gather}
\end{subequations}
where $d\equiv LK_s/A$ is the dimensionless surface-pinning parameter that relates
the exchange to the surface anisotropy and the film thickness and $\pPar\equiv \oM
L\hbar^2\gMix/4Ae^2$ is a dimensionless constant relating the exchange stiffness and
the spin-mixing conductance.

\subsection{Spin Accumulation in NM and Spin Backflow\label{sec:spin_accum}}
The pumped spin current induces a spin accumulation, $\bvec\muS=\muS\buvec s$, in
the normal metal. Here, $\buvec s$ is the spin-polarization axis, and
$\muS=(\mu_\uparrow-\mu_\downarrow)/2$ is half of the difference between chemical
potentials for spin-up and spin-down electrons in the NM.

As the spin accumulation is a direct consequence of the spin dynamics in the FI (see
Eq.~\eqref{eq:sp_current}), the spin accumulation cannot change faster than the
magnetization dynamics at the interface. Thus, assuming that spin-flip processes in
the NM are must faster than the typical precession frequency of the magnetization in
the FI\cite{PhysRevB.66.224403}, we can neglect precession of the spin accumulation
around the applied field and any decay in the NM. With this assumption, the
spin-diffusion equation
$\pderiv{\bvec\muS}{t}=D\nabla^2\bvec\muS-\frac{\bvec\muS}{\tsf}$, where $D$ is the
spin-diffusion constant, and $\tsf$ is the material-specific average spin-flip
relaxation time, becomes
\begin{equation}
    \bvec\muS\approx \lsf^2\nabla^2\bvec\muS,
    \label{eq:spin_diffusion_approx}
\end{equation}
where $\lsf\equiv\sqrt{\tsf D}$ is the average spin-flip relaxation length.

The spin accumulation results in a backflowing spin-current density, given by
\begin{equation}
    \bvec j^{(s)}_\text{bf}(L/2)=
    \frac{\hbar\gMix}{e^2M_s^2}
    \left[\Mtotal\times\Big(\Mtotal\times\bvec\muS(\bvec
    r,t)\Big)\right]_{\xi=L/2},
    \label{eq:backflow_current}
\end{equation}
where the positive sign indicates flow from the NM into the FI. This spin current
creates an additional spin-transfer torque on the magnetization at the interface
\begin{equation}
    \bvec\tau_\text{bf}=
    -\frac{\gamma\hbar\gMix}{e^2M_s^2}\delta\Big(\xi-\frac{L}{2}\Big)
        \Mtotal\times\left(\Mtotal\times\bvec\muS\right) \, .
    \label{eq:tau_backflow}
\end{equation}

Because the spin accumulation is a direct result of the pumped spin current, it must
have the same orientation as the $\Mtotal\times\partial_t{\Mtotal}$ term in
Eq.~\eqref{eq:tau_sp}. That term is comprised of two orthogonal components: the
$1^\text{st}$-order term $M_s\buvec z\times\dot{\bvec m}$ in the $xy$ plane, and the
$2^\text{nd}$-order term $\bvec m\times\dot{\bvec m}$ oriented along $\buvec z$.
Because the magnetization is a real quantity, care must be taken when evaluating the
$2^\text{nd}$-order term. Using Eq.~\eqref{eq:magn_xizeta}, the $2^\text{nd}$-order
pumped spin current is proportional to
\begin{multline}
    \Re\{\bvec m\}\times\partial_t\Re\{\bvec m\}\Big|_{\xi=L/2}=
        \eu^{-2\Im\{\omega\}t}\Re\{\omega\}\times\\
    \times\buvec z\Big[\Im X_Q\Re Y_Q-\Re X_Q\Im Y_Q\Big],
    \label{eq:mcrossmdot_real}
\end{multline}
which is a decaying \textit{direct-current} (DC) term. This is in contrast to the
$1^\text{st}$-order term, which is an \textit{alternating-current} (AC) term. Thus, we
write the spin accumulation as
\begin{equation}
    \bvec\muS=\muS_\text{AC}(\buvec z\times\hat{\bvec
    m_t})+\muS_\text{DC}\buvec z,
    \label{eq:mu_comp}
\end{equation} where we have used the shorthand notation $\bvec m_t=\dot{\bvec
m}(\xi=L/2)$, such that $\hat{\bvec m}_t=\bvec m_t/|\bvec m_t|$, which in general is
not parallel to $\bvec m$ but guaranteed to lie in the $xy$ plane. Inserting
Eq.~\eqref{eq:mu_comp} into Eq.~\eqref{eq:spin_diffusion_approx} gives one equation
each for the AC and DC components of the spin accumulation,
\begin{equation}
    \frac{\partial^2\muS_j}{\partial\xi^2}=\lsfj^{-2}\muS_j,
    \label{eq:spin_diffusion_comp}
\end{equation}
where $j$ denotes either the AC or DC case and $l_\text{sf,DC}=\lsf$ while
$l_\text{sf,AC}=\lsf(1+\lsf^2Q^2)^{-1/2}$ because $\bvec m_t\propto\exp (\iu(\omega
t-Q\zeta))$. Eq.~\eqref{eq:spin_diffusion_comp} can be solved by demanding
spin-current conservation at the NM boundaries: at the free surface of the NM, there
can be no crossing spin current; thus, the $\xi$ component of the spin-current
density must vanish there, $\partial_\xi\muS_j|_{\xi=L/2+d}=0$. Similarly, by
applying conservation of angular momentum at the FI-NM interface, the net
spin-current density crossing the interface, due to spin pumping and backflow, must
equal the spin current in the NM layer, giving
\begin{multline}
    \left[
        -\frac{\hbar^2\gMix}{2e^2M_s^2}
            \bvec M\times\pderiv{\bvec M}{t}
        +\frac{\hbar\gMix}{e^2M_s^2}
            \bvec M\times\Big(\bvec M\times\bvec\muS\Big)
    \right]_{\xi=L/2} \\
    = -\frac{\hbar\sigma}{2e^2}\partial_\xi\bvec\muS|_{\xi=L/2},
    \label{eq:spin_balance}
\end{multline}
where $\sigma$ is the conductivity of the NM. Using these boundary conditions, we
recover the solutions (see, e.g., \cite{PhysRevB.66.224403,PhysRevLett.110.217602})
\begin{equation}
    \muS_j=\muS_{j,0}
        \frac{\sinh\left(\lsfj^{-1}\big[\xi-(L/2+d)\big]\right)}
             {\sinh\left(-\frac{d}{\lsfj}\right)},
\end{equation}
where $\muS_{j,0}$ is time dependent, and depends on the $\zeta$ coordinate only in
the AC case. We find that the AC and DC spin accumulations $\muS_{j,0}$ are given by
\begin{gather}
    \muS_{\text{AC},0}=-\frac{\hbar}{2}\frac{m_t}{M_s}
        \left[
            1+\frac{\sigma}{2\gMix\lsfAC}\coth\left(\frac{d}{\lsfAC}\right)
        \right]^{-1},\label{eq:mu0_AC}\\
    \begin{aligned}
    \muS_{\text{DC},0}=&
            -\frac{\lsf\hbar}{\sigma M_s^2}\gMixTilde
        \tanh\left(\frac{d}{\lsf}\right)
            \buvec z\cdot\left[\bvec m\times\dot{\bvec m}\right]_{\xi=L/2},
        % OLD explicit expression:
            %\eu^{-2\Im\{\omega\}t}\Re\{\omega\}\times\\
            %{}&\times\frac{\iu}{2}\Big[Y_Q(\xi)X_Q^*(\xi)-X_Q(\xi)Y_Q^*(\xi)\Big]_{\xi=L/2}
    \end{aligned}\label{eq:mu0_DC}
\end{gather}
where $\gMixTilde$ is a renormalized mixing conductance, which is given by
\begin{equation}
    \gMixTilde = \gMix\left\{
        1-\left[
            1+\frac{\sigma}{2\gMix\lsfAC}\coth\left(\frac{d}{\lsfAC}\right)
        \right]^{-1}\right\}.
\end{equation}
This scaling of $\gMix$ occurring in the DC spin accumulation originates from the
second-order spin backflow due to the AC spin accumulation that is generated in the
normal metal.

Adding both the spin-pumping and the backflow torques to Eq.~\eqref{eq:LLG} and
repeating the linearization procedure from Sec.~\ref{sec:linearization}, we find that
the AC spin accumulation renormalizes the pure spin-mixing conductance. Thus,
the addition of the backflow torque can be accounted for by replacing $\gMix$ with
$\gMixTilde$ in the boundary conditions of Eqs.~\eqref{eq:BClin}, making the boundary
conditions $Q$-dependent in the process.
%--------------------------------------------------------------------
\begin{table}
    \caption{\label{tab:parameters}Typical values for the parameters used in the
    calculations.\cite{PhysRevLett.107.066604, 2013arXiv1302.6697J,
    2013arXiv1302.7091Q, rezende:012402, Hillebrands.gMix.2013,
    PhysRevLett.112.197201}}
    \begin{ruledtabular}
    \begin{tabular}{ccc}
    Parameter & Value & Unit \\
    \hline
    $A$ & $3.66\cdot10^{-7}$ & $\unit{erg\,cm^{-1}}$ \\ 
    $\alpha$ & $3\cdot10^{-4}$& --\\
    $K_s$ & $0.05$ & $\unit{erg\,cm^{-2}}$ \\
    $\gMix$ & $8.18\cdot10^{22}$ & $\unit{cm^{-1}\,s^{-1}}$ \\
    $\gamma$ & $1.76\cdot10^{7}$ & $\unit{G^{-1}\,s^{-1}}$ \\
    $4\pi M_s$ & $1750$ & $\unit{G}$ \\
    $\sigma$\cite{} & $8.45\cdot 10^{16}$ & $\unit{s^{-1}}$ \\
    $d$ & 50 & $\unit{nm}$\\
    $\lsf$ & 7.7 & $\unit{nm}$\\
    $\Theta$ & 0.1 & --\\
    \end{tabular}
    \end{ruledtabular}
\end{table}
%--------------------------------------------------------------------
Using the values from Table~\ref{tab:parameters}, which are based on typical values
for a YIG-Pt bilayer system, we obtain $\gMixTilde/\gMix\sim0.4$ for $QL\ll 1$,
whereas $\gMixTilde/\gMix\to1$ for large values of $QL$. Thus, AC backflow is
significant for long-wavelength modes and should be considered when estimating
$\gMix$ from the linewidth broadening in \textit{ferromagnetic resonance} (FMR)
experiments.\cite{Hillebrands.gMix.2013}
%--------------------------------------------------------------------

\subsection*{Inverse Spin Hall Effect}
%--------------------------------------------------------------------
The \textit{inverse spin Hall effect} (ISHE) converts a spin current in the NM to an
electric potential through the spin-orbit coupling in the NM. For a spin current in
the $\buvec\xi$ direction, the ISHE electric field in the NM layer is $\bvec
E_\text{ISHE} = -e^{-1}\Theta\langle(\partial_\xi\bvec\muS)
\times\buvec\xi\rangle_\xi$, where $\Theta$ is the dimensionless spin-Hall angle, and
$\langle\cdot\rangle_\xi$ is a spatial average across the NM layer, i.e., for
$\xi\in(L/2,L/2+d)$. Using the previously calculated spin accumulation, we find that
the AC electric field is
\begin{align}
    \bvec E_\text{ISHE}^\text{AC}=&
        -\Theta\frac{\hbar}{2deM_s}\left[
            1+\frac{\sigma}{2\gMix\lsfAC}\coth\left(\frac{d}{\lsfAC}\right)
        \right]^{-1}\times \nonumber\\
    {}&    \quad\times\big[
            -\buvec\eta(-m_{t,y}\cos\theta\cos\phi+m_{t,x}\sin\phi)+\nonumber\\
    {}& \qquad+\buvec\zeta(-m_{t,x}\cos\phi-m_{t,y}\cos\theta\sin\phi)\big],
    \label{eq:Eishe_ac}
\end{align}
where
\begin{equation}
    m_{t,i}=-[\Im{\omega}\Re{m_i}+\Re{\omega}\Im{m_i}]_{\xi=L/2},
\end{equation}
and $i=x,y$. For BVMSW ($\theta=\pi/2, \phi=0$) modes, the AC field points along
$\buvec\zeta$, whereas for MSSW ($\theta=\phi=\pi/2$) modes, it points along
$\buvec\eta$ (i.e., in plane, but transverse to $\zeta$; see
Fig.~\ref{fig:geometry}). Notice that for both BVMSW and MSSW mode geometries, only
the $x$ component of $\bvec m_t$ contributes to the field. In contrast, for FVMSW
($\theta=0$) modes, the field points somewhere in the $\eta\zeta$ plane, depending on
the ratio of $m_{t,x}$ to $m_{t,y}$.

Similarly to the AC field, the DC ISHE electric field is given by
\begin{align}
    \bvec E_\text{ISHE}^\text{DC}=
        \Theta\frac{\muS_{\text{DC},0}}{de}
            \sin\theta(\buvec\eta\cos\phi-\buvec\zeta\sin\phi),
    \label{eq:Eishe_dc}
\end{align}
which is perpendicular to the AC electric field and zero for the FVMSW mode
geometry.

The total time-averaged energy in the ferromagnet $\mathcal{E}_\text{total}$ (see
\citet{morgenthaler1972dynamic}) is given by
\begin{equation}
    \langle\mathcal{E}_\text{total}\rangle_T=\int_\text{ferrite}\Re\left[
                -\iu\pi\frac{\omega^*}{\oM}(\bvec m\times\bvec m^*)\buvec z
            \right]\,\dd V,
    \label{eq:EtotalFM}
\end{equation}
where the integral is taken over the volume of the ferromagnet.

Because the DC ISHE field is in-plane, the voltage measured per unit distance along
the field direction, $\buvec\Lambda=\buvec\eta\cos\phi-\buvec\zeta\sin\phi$, can be
used to construct an estimate of the mode efficiency. Taking the one-period time
average of Eq.~\eqref{eq:Eishe_dc} using Eq.~\eqref{eq:mu0_DC} and normalizing it by
Eq.~\eqref{eq:EtotalFM} divided by the in-plane surface area, $\mathcal{A}$, we find
an amplitude-independent measure of the DC ISHE:
\begin{multline}
    \epsilon^\text{DC}=\frac{\langle e\buvec \Lambda\cdot\bvec E_\text{ISHE}^\text{DC}\rangle_T}
        {\langle\mathcal{E}_\text{total}\rangle_T/\mathcal{A}}=
            -2\gamma\Theta\frac{\lsf\hbar}{d \sigma M_s}\gMixTilde
            \tanh\left(\frac{d}{\lsf}\right)\sin\theta\times\\
        \times\frac{
            \Re\left[
                -\iu\frac{\omega^*}{\oM}(\bvec m\times\bvec m^*)\buvec z
            \right]_{\xi=L/2}
        }{
            \int_{-L/2}^{L/2}\Re\left[
                -\iu\frac{\omega^*}{\oM}(\bvec m\times\bvec m^*)\buvec z
            \right]\,\dd\xi
        }
    \label{eq:Eishe_dc_normalized},
\end{multline}
given in units of $\unit{cm}$, and where $\{\cdot\}^*$ denotes complex conjugation. 

Similarly, the AC ISHE electric field, being time-varying, will contribute a power
density that, when normalized by the power density in the ferromagnet, becomes 
\begin{multline}
    \epsilon^\text{AC}=\frac{\langle\sigma\big(\bvec E_\text{ISHE}^\text{AC}\big)^2\rangle_T}
         {\frac{\Re\{\omega\}}{2\pi\mathcal{A}L}
            \langle\mathcal{E}_\text{total}\rangle_T}=
    %\frac{\Re\{\frac{\sigma}{2}|\bvec E|^2\}}
    %     {\langle\mathcal{E}_\text{total}\rangle_T/\mathcal{A}}=
    \frac{\pi\sigma}{\Re\{\omega\}}\left(\frac{\Theta\hbar}{2deM_s}\right)^2\times\\
    \times\left[
        1+\frac{\sigma}{2\gMix\lsfAC}\coth\left(\frac{d}{\lsfAC}\right)
    \right]^{-2}\times\\
    \times\frac{|m_{t,x}|^2+\cos^2\theta|m_{t,y}|^2
    }{
        \frac{1}{L}\int_{-L/2}^{L/2}\Re\left[
            -\iu\frac{\omega^*}{\oM}(\bvec m\times\bvec m^*)\buvec z
        \right]\,\dd\xi
    }\ .
    \label{eq:Eishe_ac_normalized}
\end{multline}
To be able to calculate explicit realizations of the mode-dependent equations
Eqs.~\eqref{eq:Eishe_dc_normalized} and \eqref{eq:Eishe_ac_normalized}, one will need
to first calculate the dispersion relation and mode profiles in the ferromagnet.

\section{Spin-Pumping Theory for Travelling Spin Waves\label{sec:sptheory}}
%-------------------------------------------------------------------------------
Because, the linearized boundary conditions (see Eqs.~\eqref{eq:BClin}) explicitly
depend on the eigenfrequency $\omega$, we cannot apply the method of expansion in the
set of pure exchange spin waves, as was performed by \citet{Kalinikos1986}. Instead, we
analyze and solve the system directly for small values of $QL$, whereas the
dipole-dipole regime of $QL\sim 1$ is explored using numerical computations in
Sec.~\ref{sec:numerics}.

\subsection{Long-Wavelength Magnetostatic Modes\label{sec:long_wavelength_modes}}
%-------------------------------------------------------------------------------
When $QL\ll 1$ Eq.~\eqref{eq:motion} is simplified to
\begin{multline}
    \left\{
    \begin{pmatrix}
        \sin^2\theta & 0 \\
        0 & 0
    \end{pmatrix}
    +\iu\frac{\omega}{\oM}
    \begin{pmatrix}
        \alpha & -1 \\
        1 & \alpha
    \end{pmatrix}+\right.\\
    \left.+\openone\left[
        \frac{\oH}{\oM}-8\pi\frac{\gamma^2A}{\oM^2}\deriv[2]{}{\xi}
    \right]\right\}
    \cdot\dynMQ(\xi)=0,
    \label{eq:motion_QL_small}
\end{multline}
where the $1^\text{st}$-order matrix term describe the dipole-induced shape
anisotropy and stems from $\widehat{\mathcal{G}}_{xy}$ (see \cite{Kalinikos1986}). We
make the ansatz that the magnetization vector in Eq.~\eqref{eq:magn_xizeta} is
composed of plane waves, e.g., $\bvec m_Q(\xi)\propto \eu^{\iu k\xi}$. Inserting this
ansatz into Eq.~\eqref{eq:motion_QL_small} produces the dispersion relation
\begin{multline}
    \Big(\ooM\Big)^2 =
    \Big(\frac{\oH}{\oM}+\lEx^2k^2+\iu\alpha\ooM\Big)\times\\
        \quad\times\Big(\frac{\oH}{\oM}+\lEx^2k^2+\sin^2\theta+\iu\alpha\ooM\Big),
    \label{eq:bulk_dispersion_squared}
\end{multline}
where $\lEx\equiv\lExExplicitInline$ is the \textit{exchange length}. Keeping only
terms to first order in the small parameter $\alpha$, we arrive at 
\begin{align}
    \frac{\omega(k)}{\oM}={}&
        \pm\sqrt{
            \Big(\oHoM+\lEx^2k^2\Big)
            \Big(\oHoM+\lEx^2k^2+\sin^2\theta\Big)}+\nonumber\\
    {}&\quad
        +\iu\alpha\Big(\oHoM+\lEx^2k^2+\frac{\sin^2\theta}{2}\Big).
    \label{eq:bulk_dispersion}
\end{align}
The boundary conditions in Eq.~\eqref{eq:BClin} depend explicitly on $\omega$ and $k$
and give another equation $k=k(\omega)$ to be solved simultaneously with
Eq.~\eqref{eq:bulk_dispersion}. However, in the absence of spin pumping, i.e., when
the spin-mixing conductance vanishes $\gMix\to0$, it is sufficient to insert the
constant $k$ solutions from the boundary conditions into
Eq.~\eqref{eq:bulk_dispersion} to find the eigenfrequencies.

Different wave vectors can give the same eigenfrequency. It turns out that this is
possible when $\omega(k)=\omega(\iu\kappa)$, which has a non-trivial solution relating
$\kappa$ to $k$:
\begin{align}
    \lEx^2\kappa^2 =
    {}& \sin^2\theta+\lEx^2k^2+2\oHoM \pm\iu2\alpha\omega(k)/\oM.
%    \iu2\alpha\sqrt{\Big(\oHoM+\lEx^2k^2\Big)\Big(\oHoM+\lEx^2k^2+\sin^2\theta\Big)}.
    \label{eq:kappa_k}
\end{align}

With these findings, a general form of the magnetization is 
\begin{align}
    \dynMQ(\xi)
        ={}&\begin{pmatrix}1\\r(k)\end{pmatrix}\Big[
             C_1\cos\big(k(\xi+\frac{L}{2})\big)
            +C_2\sin\big(k(\xi+\frac{L}{2})\big)\Big]+\nonumber\\
        {}&+\begin{pmatrix}1\\r(\iu\kappa)\end{pmatrix}\Big[
             C_3\cosh\big(\kappa(\xi+\frac{L}{2})\big)
            +C_4\sinh\big(\kappa(\xi+\frac{L}{2})\big)\Big],
    \label{eq:magn_ansatz}
\end{align}
where ${\{C_i\}}$ are complex coefficients to be determined from the boundary
conditions, and where $\kappa=\kappa(k)$ is given by Eq.~\eqref{eq:kappa_k}. The
ratio between the transverse components of the magnetization, $r(k)=Y_Q/X_Q$, is
determined from the bulk equation of motion (see Eq.~\eqref{eq:motion_QL_small}) and
is in linearized form
\begin{align}
    r(k) ={}& -\frac{\alpha\sin^2\theta\pm2\iu\sqrt{
                \Big(\oHoM+\lEx^2k^2\Big)\Big(\oHoM+\lEx^2k^2+\sin^2\theta\Big)}
            }{2\Big(\oHoM+\lEx^2k^2\Big)},
\end{align}
implying elliptical polarization of $\dynMQ$ when $\theta\neq 0$.

Inserting Eq.~\eqref{eq:magn_ansatz} into Eq.~\eqref{eq:exchange_BC_lower} only leads
to a solution when $k=0$, such that $C_2=C_4=0$ in the general case. By solving
Eq.~\eqref{eq:BClinY} for $C_3$, we find
\begin{multline}
 \frac{C_3}{C_1} =
     -\frac{\oHoM+\lEx^2k^2+\sin^2\theta+\iu\alpha\ooM}
      {\oHoM-\lEx^2\kappa^2+\sin^2\theta+\iu\alpha\ooM}\times\\
    \times\frac{(\iu\ooM\pParTilde+d\cos^2\theta)\cos(kL)-kL\sin(kL)}
        {(\iu\ooM\pParTilde+d\cos^2\theta)\cosh(\kappa L)+\kappa L\sinh(\kappa L)},
    \label{eq:C3}
\end{multline}
where $\pParTilde\equiv\pPar|_{\gMix\to\gMixTilde}$ is the pumping parameter altered
by the AC spin backflow from the NM (see Section~\ref{sec:spin_accum}). $C_1$ is
chosen to be the free parameter that parameterizes the dynamic magnetization
amplitude, which can be determined given a particular excitation scheme.
Linearization of Eq.~\eqref{eq:C3} with respect to $\alpha$ is straightforward, but
the expression is lengthy; we will therefore not show it here.

Inserting the ansatz with $C_2=C_4=0$ and $C_3$ given by Eq.~\eqref{eq:C3} into
Eq.~\eqref{eq:BClinX} gives the second equation for $k$ and $\omega$ (the first is
Eq.~\eqref{eq:bulk_dispersion}). In the general case, the number of terms in this
equation is very large; thus, we describe it as
\begin{equation}
    f(k,\omega,\alpha,\tilde\pPar)=0,
    \label{eq:f_boundary}
\end{equation}
i.e., an equation that depends on the wave vector $k$, frequency $\omega$, Gilbert
damping constant $\alpha$ and spin-pumping parameter $\tilde\pPar$.

Because both the bulk and interface-induced dissipation are weak, $\alpha\ll1$,
$\pParTilde\ll1$, the wavevector is only slightly perturbed with respect to a system
without dissipation, i.e., $k\to k+\dk$ where $\lEx\dk\ll1$. It is therefore
sufficient to expand $f$ up to $1^\text{st}$ order in these small quantities:
\begin{multline}
    f(k,\omega,0,0)+\left.(\pParTilde)\pderiv{f}{\pParTilde}\right|_0
        +\left.\alpha\pderiv{f}{\alpha}\right|_0+\\
    +(\lEx\dk)\left.\pderiv{f}{(\lEx\dk)}\right|_0 \approx 0,
\end{multline}
where the sub-index 0 means evaluation in a system without dissipation, i.e., when
$(\alpha,\pParTilde,\dk)=(0,0,0)$. By solving the system of equations in the absence
of dissipation, $f(k,\omega,0,0)=0$, the dissipation-induced change in the wave
vector $\dk$ is given by
\begin{equation}
\dk\approx
-\frac{\pParTilde\left.\pderiv{f}{\pParTilde}\right|_0+
        \alpha\left.\pderiv{f}{\alpha}\right|_0}
    {\lEx \left.\pderiv{f}{(\lEx\dk)}\right|_0}.
    \label{eq:dk}
\end{equation}
In turn, this change in the wave vector should be inserted into the dispersion
relation of Eq.~\eqref{eq:bulk_dispersion_squared} to find the dissipation.
Inspecting Eq.~\eqref{eq:bulk_dispersion_squared}, we note that $\delta k$-induced
additional terms proportional to $\omega$ are of the form $(k+\dk)^2-k^2\approx 2 k
\delta k$ which renormalize the Gilbert-damping term $\iu\alpha\ooM$. Thus, in
Eq.~\eqref{eq:dk}, there are terms proportional to the frequency in both terms in the
numerator. We extract these terms $\propto\iu\ooM$ by differentiating with respect to
$\omega$ and define the renormalization of the Gilbert damping, i.e.,
$\alpha\to\alpha+\Delta\alpha$, from spin pumping as
\begin{align}
\Delta\alpha = {}& \frac{\iu 2\lEx k
\oM\partial_{\omega}\big(\lEx\dk|_{\alpha=0}\big)}
    {\iu 2\lEx k\oM\partial_\omega\big(\lEx\dk|_{\pParTilde=0}\big)-1}\,,
    \label{eq:deltaAlphaDef}
\end{align}
where $\partial_\omega$ represents the derivative with respect to $\omega$ and $k$
is the solution to the $0^\text{th}$-order equation. Note that in performing a further
local analysis around some point $k_0$ in the $k$-space of Eq.~\eqref{eq:f_boundary}, a
series expansion of $f$ around $k_0$ must be performed before evaluating
Eqs.~\eqref{eq:dk} and \eqref{eq:deltaAlphaDef}.

Eq.~\eqref{eq:deltaAlphaDef} is generally valid, except when $d=0$ and $kL\to0$,
which we discuss below. In the following section, we will determine explicit
solutions of the $0^\text{th}$-order equation for some key cases, and mapping out the
spin-wave dispersion relations and dissipation in the process.

\subsection{No Surface Anisotropy (\texorpdfstring{$d=0$}{d=0})
\label{seq:long_wavelength_modes_no_surf}}
%-------------------------------------------------------------------------------
Let us first investigate the case of a vanishing surface anisotropy. In this case,
the $0^\text{th}$-order expansion of Eq.~\eqref{eq:f_boundary} has a simple form and is
independent of the magnetization angle $\theta$. The equation to determine $k$ is
given by
\begin{equation}
    kL\tan(kL)=0,
    \label{eq:BC_noPinning}
\end{equation}
with solutions $k=n\pi/L$, where $n\in\mathbb{Z}$. Similarly, the expression for $\dk$
is greatly simplified, $\dk_n= \iu\ooM\frac{\pParTilde}{n\pi} \frac{\lEx}{L}$, $n\neq
0$, such that the mode-dependent Gilbert damping is 
\begin{equation}
    \Delta\alpha_n = 2\pParTilde\left(\frac{\lEx}{L}\right)^2,\quad n\neq 0\,.
    \label{eq:deltaAlpha_n}
\end{equation}
For the macrospin mode, when $n=0$, the linear expansion in $\dk$ becomes
insufficient. This is because $kL\tan(kL)\sim(kL)^2$ for $kL\to0$; thus, we must
expand the function $f$ to second order in the deviation $\dk$ around $kL=0$. For
$d=0$, we find that the boundary condition becomes
$\dk^2L^2=\iu\ooM\pParTilde\lEx^2$, and when inserted into
Eq.~\eqref{eq:bulk_dispersion_squared}, it immediately gives
\begin{equation}
    %\Delta\alpha_0=8\pi\frac{\gamma^2A\chi}{L^2\oM}=\frac{\gamma\hbar^2\gMix}{2e^2M_sL},
    \Delta\alpha_0=\pParTilde\left(\frac{\lEx}{L}\right)^2=\frac{1}{2}\Delta\alpha_n,
    \label{eq:deltaAlpha_0}
\end{equation}
which is the macrospin renormalization factor found in Ref.~\onlinecite{PhysRevLett.88.117601}.
Using a different approach, our results in this section reproduce our previous result
that the renormalization of the Gilbert damping for standing waves is twice the
renormalization of the Gilbert damping of the macrospin.\cite{PhysRevLett.111.097602}
Next, we will obtain analytical results beyond the description in Ref.
\onlinecite{PhysRevLett.111.097602} for the enhancement of the Gilbert damping in the
presence of surface anisotropy.

\subsection{Including Surface Anisotropy
(\texorpdfstring{$d\neq0$}{d!=0})}\label{sec:SA_incl}
%-------------------------------------------------------------------------------
In the presence of surface anisotropy, the out-of-plane and in-plane field
configurations must be treated separately. This distinction is because the boundary
condition Eq.~\eqref{eq:f_boundary} has different forms for the two configurations in
this scenario.

\subsubsection{Out-of-plane Magnetization\label{sec:long_wl_oop}}
%-------------------------------------------------------------------------------
When the magnetization is out of plane, i.e., $\theta=0$, the spin-wave excitations
are circular and have a high degree of symmetry. A simplification in this geometry is
that the coefficient $C_3=0$. In the absence of dissipation, the boundary
condition Eq.~\eqref{eq:f_boundary} determining the wave vectors becomes
\begin{equation}
    kL\tan(kL)=d.
    \label{eq:BClin_theta_zero}
\end{equation}
Let us consider the effects of the two different anisotropies in this geometry.

\paragraph{Easy-Axis Surface Anisotropy ($d>0$):\label{sec:EA_perp}} When $d\sim1$ or
larger, the solutions of Eq.~\eqref{eq:BClin_theta_zero} are displaced from the zeroes
of $\tan(kL)$, i.e., the solutions we found in the case of no surface anisotropy, and
towards the upper poles located at $k_uL=(2n+1)\pi/2$, where $n=0,1,2,\ldots$. We
therefore expand $f$ in Eq.~\eqref{eq:f_boundary} (and thus also in
Eq.~\eqref{eq:BClin_theta_zero}) into a Laurent series around the poles from the
first negative order up to the first positive order in $kL$ to solve the boundary
condition for $kL$, giving
%-------------------------------------------
\newcommand{\kupL}{k_uL} % read as "k upper pole" * L
%-------------------------------------------
\begin{align}
    kL\approx {}&
        \frac{\lEx}{L}\frac{3(1+d)+2(\kupL)^2-\sqrt{12(\kupL)^2+9(1+d)^2}}{2\kupL}.
    \label{eq:kL_EA_oop}
\end{align}
%-------------------------------------------
\begin{figure}
    \centering
    \includegraphics[width=\linewidth]{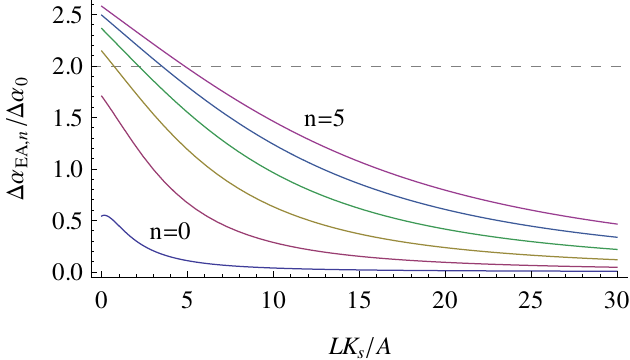}
    \caption{The ratio of enhanced Gilbert damping
    $\Delta\alpha_\text{EA,n}/\Delta\alpha_0$ in a system with easy-axis surface
    anisotropy versus the enhanced Gilbert damping of macrospin modes in systems with
    no surface anisotropy as a function of surface-anisotropy energy. $n$ refers to
    the mode number, where $n=0$ is the uniform-like mode. The dashed line represents
    the ratio $\Delta\alpha_n/\Delta\alpha_0$ in the case of no surface anisotropy
    (see Eq.~\eqref{eq:deltaAlpha_n}). \label{fig:EA_dissipation_factor}}
\end{figure}
%-------------------------------------------
Using this result and the Laurent-series expansion for $f$ in Eq.~\eqref{eq:dk} and
Eq.~\eqref{eq:deltaAlphaDef}, we find the Gilbert-damping renormalization term
($\alpha\to\alpha+\Delta\alpha^\text{(oop)}_\text{EA,n}$) and the ratio between the
modes
\begin{align}
    \frac{\Delta\alpha^\text{(oop)}_\text{EA,n}}{\Delta\alpha_0}\approx& 
            3\big(3(1+d)+2(\kupL)^2-\sqrt{12(\kupL)^2+9(1+d)^2}\big)\times\nonumber\\
        {}&\quad\times\frac{
                \big(\sqrt{4(\kupL)^2+3(1+d)^2}-\sqrt{3}(1+d)\big)
            }{ 2(\kupL)^2\sqrt{4(\kupL)^2+3(1+d)^2} }.
    \label{eq:deltaAlpha_EA_perp}
\end{align}
This ratio is plotted in Figure \ref{fig:EA_dissipation_factor} for $n\le 5$. We see
that the ratio vanishes for large values of $d$. For small values of the anisotropy
energy $d$, the approximate ratio exceeds the exact result of the ratio we found in
the limiting case of no surface anisotropy (see Eq.~\eqref{eq:deltaAlpha_n}). For
moderate values of $d\sim 5$, the expansion around the upper poles is sufficient, but only
for the first few modes. This implies that moderate-strength easy-axis surface
anisotropy quenches spin pumping for the lowest excited modes but does not affect
modes with higher transverse exchange energy.

\paragraph{Easy-Plane Surface Anisotropy ($d<0$):\label{sec:oop_EP}}
Easy-plane surface anisotropy is represented by a negative surface anisotropy $d$ in
Eq.~\eqref{eq:BClin_theta_zero}. In this case, the boundary condition must be treated
separately for the uniform-like ($n=0$) mode and the higher excitations. When
$|d|>1$, we can obtain a solution by expanding along the imaginary axis of $kL$. This
corresponds to expressing the boundary condition in the form $-\iu kL\tanh(\iu
kL)=-|d|$, with the asymptotic behavior $kL\approx -\iu|d|$. Using the asymptotic
form of the boundary condition in Eqs.~\eqref{eq:dk} and calculating the
renormalization of the Gilbert damping using Eq.~\eqref{eq:deltaAlphaDef}, we find
that the renormalization is $\alpha\to\alpha+\Delta\alpha^\text{(oop)}_\text{EP,0}$,
where
\begin{equation}
    \frac{\Delta\alpha^\text{(oop)}_\text{EP,0}}{\Delta\alpha_0}=2|d|.
    \label{eq:deltaAlpha_FVMSW_EP_0}
\end{equation}
Thus, the Gilbert damping of the lowest mode is much enhanced by increasing surface
anisotropy. The surface-anisotropy mode is localized at the surface because it decays
from the spin-active interface and into the film. Because the effective volume of the
mode is reduced, spin pumping more strongly causes dissipation out of the mode and
into the normal metal.

%-------------------------------------------
\newcommand{\klpL}{k^{(l)}_nL} % read as "k lower pole" * L
%-------------------------------------------
For the higher modes ($n>0$), the negative term on the rhs of
Eq.~\eqref{eq:BClin_theta_zero} forces the $kL$ solutions closer to the negative,
lower poles of $\tan(kL)$, located at $\klpL=(2n-1)\pi/2$, where $n=1,2,3,\ldots$. We
repeat the procedure used for the EA case by expanding $f$ into a Laurent series
around these lower poles, arriving at
\begin{align}
    kL\approx {}&
        \frac{3(1-|d|)+2(\klpL)^2+\sqrt{12(\klpL)^2+9(1-|d|)^2}}{2\klpL}.
    \label{eq:kL_EP_oop}
\end{align}
Using this relation and the new lower-pole Laurent expansion for $f$,
Eqs.~\eqref{eq:dk} and \eqref{eq:deltaAlphaDef} give us the renormalization of the
Gilbert damping ($\alpha\to\alpha+\Delta\alpha^\text{(oop)}_\text{EP,n}$) and the
ratio
%-------------------------------------------
\begin{figure}
    \centering
    \includegraphics[width=\linewidth]{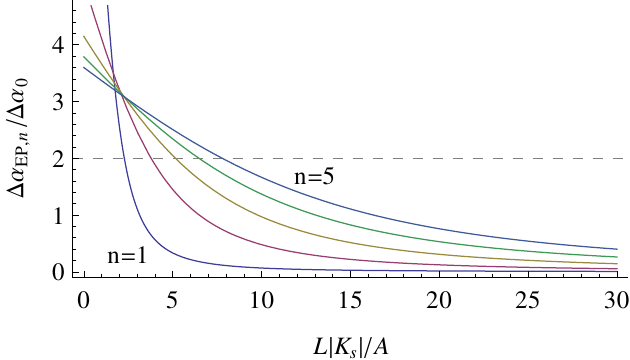}
    \caption{Plot of $\Delta\alpha^\text{(oop)}_\text{EP,n}/\Delta\alpha_0$.
    The dashed line represents the ratio $\Delta\alpha_n/\Delta\alpha_0$ in the case
    of no surface anisotropy (see
    Eq.~\eqref{eq:deltaAlpha_n}).\label{fig:EP_dissipation_factor}}
\end{figure}
%-------------------------------------------
\begin{align}
    \frac{\Delta\alpha^\text{(oop)}_\text{EP,n}}{\Delta\alpha_0}\approx&
        3\big(3(1-|d|)+2(\kupL)^2+\sqrt{12(\kupL)^2+9(1-|d|)^2}\big)\times\nonumber\\
        {}&\quad\times\frac{
                \big(\sqrt{4(\kupL)^2+3(1-|d|)^2}+\sqrt{3}(1-|d|)\big)
            }{ 2(\kupL)^2\sqrt{4(\kupL)^2+3(1-|d|)^2} }.
    \label{eq:deltaAlpha_EP_perp}
\end{align}
This ratio is plotted in Figure \ref{fig:EP_dissipation_factor} from $n=1$ up to
$n=5$. We see that the ratio vanishes for large values of $|d|$. Similar to the case
of EA surface anisotropy, the approximation breaks down for large $n$ and/or small
values of $|d|$.

Whereas the $n=0$ mode exhibits a strong spin-pumping enhanced dissipation in this
field configuration, the DC ISHE field vanishes when $\theta=0$ (see
Eq.~\eqref{eq:Eishe_dc}). This is one of the reasons why this configuration is seldom
used in experiments. However, this configuration can lead to a significant AC ISHE,
and a similar AC signal was recently detected\cite{PhysRevLett.111.217204}. Because
of the strong dissipation enhancement, the EP surface anisotropy induced localized
mode in perpendicular magnetization geometry could be important in future
experimental work.

\subsubsection{In-plane Magnetization\label{sec:long_wl_inplane_SA}}
We will now complete the discussion of the spin-pumping enhanced Gilbert damping by
treating the case in which the magnetization is in plane ($\theta=\pi/2$). For such
systems, the coefficient $C_3\neq0$, and the $0^\text{th}$-order expansion of
Eq.~\eqref{eq:f_boundary} becomes
\begin{widetext}
\begin{equation}
 kL\tan{kL}=
     -\frac{
        d\big((\lEx k)^2+\oHoM\big)\sqrt{1+(\lEx k)^2+2\oHoM}
    }{
        \sqrt{1+(\lEx k)^2+2\oHoM}\big(1+2(\lEx k)^2+2\oHoM\big)
        -d\frac{\lEx}{L}\big(1+(\lEx k)^2+\oHoM\big)
            \coth\left(\frac{L}{\lEx}\sqrt{1+(\lEx k)^2+2\oHoM}\right)
    }.
    \label{eq:f0_inplane}
\end{equation}
\end{widetext}
For typical film thicknesses, of some hundred nanometers, we have $L/\lEx\gg1$ and
$(\lEx k)^2\ll1$ for the lowest eigenmodes. Thus, we take the asymptotic
$\coth\sim 1$ and neglect the $(\lEx k)^2$ terms, ridding the rhs of
Eq.~\eqref{eq:f0_inplane} of any $k$ dependence. Eq~\eqref{eq:f0_inplane} now becomes
similar to the out-of-plane case
\begin{equation}
    kL\tan(kL)=\dEff,
    \label{eq:BClin_inplane}
\end{equation}
where
\begin{equation}
     \dEff=-\frac{d\oHoM\sqrt{1+2\oHoM}
                }{\big(1+2\oHoM\big)^{3/2}-d\frac{\lEx}{L}\big(1+\oHoM\big)}.
    \label{eq:dEff}
\end{equation}
$\dEff$ is positive if $d<0$ and negative for $d>0$ up to a critical value
$d\lEx/L=\lEx K_s/A=\big(1+2\oHoM\big)^{3/2}/\big(1+\oHoM\big)$, where the
denominator becomes zero. For negative $d$, $|\dEff|<|d|$, whereas for positive $d$,
$|\dEff|$ is initially smaller than that of $|d|$ but quickly approaches the critical
value. With the value $K_s$ from Tab.~\ref{tab:parameters}, we have $|\dEff|<|d|$,
independent of the sign of $d$.

With this relation, we can calculate an approximate Gilbert damping renormalization
in both the EA and EP cases using the EP and EA relations, respectively, obtained in
the out-of-plane configuration. Thus,
\begin{gather}
    \Delta\alpha^\text{ip}_\text{EA,0} \approx
        \Delta\alpha^\text{oop}_\text{EP,0}|_{d\to\dEff} = 2|\dEff|, \\
    \Delta\alpha^\text{ip}_\text{EA,n} \approx
        \Delta\alpha^\text{oop}_\text{EP,n}|_{d\to\dEff},\\
    \Delta\alpha^\text{ip}_\text{EP,n} \approx
        \Delta\alpha^\text{oop}_\text{EA,n}|_{d\to\dEff}.
\end{gather}

To summarize this section regarding the enhancement of Gilbert damping, we see that the
enhancement can be very strong for the surface modes because their effective sizes
are smaller than the thickness of the film. For all other modes, the enhancement
decreases with increasing magnitude of the surface-anisotropy energy.

\section{Numerical Calculations\label{sec:numerics}}
The first step in the numerical method is to approximate the equation of motion of
Eq.~\eqref{eq:motion} into by finite-size matrix eigenvalue problem. We discretize the
transverse coordinate $\xi$ on the interval $[-L/2,L/2]$ into $N$ points labeled by
$j=1,2,\ldots,N$, and characterize the transverse discrete solutions of the dynamic
magnetization vectors $\dynMQ$ by $(m_{x,j},m_{y,j})$ of size $2N$.

We approximate the $2^\text{nd}$-order derivative arising from the exchange
interaction using a $n^\text{th}$-order central difference method. For the $n-2$
discretized points next to the boundaries, we also use $n^\text{th}$-order methods,
using forward (backward) difference schemes for the lower (upper) film boundary.
This strategy avoids the introduction of ``ghost'' points outside the interval
$[-L/2,L/2]$ to satisfy the boundary conditions.

Thus, the total operator acting on the magnetization on the left-hand side of
Eq.~\eqref{eq:motion} becomes a sparse $2N\times2N$ matrix operator. On the
right-hand side of Eq.~\eqref{eq:motion}, we also represent the convolution integral
as a $2N\times2N$ dense matrix operator, where each row is weighted according to the
extended integration formulas for closed integrals to $n$th order\cite{NR3rdEdition}.
The four $N\times N$ sub-blocks of this integration operator correspond to the four
tensor elements of $\gTensorXY$. In the final discrete form, we obtained a
$2N\times2N$ $\omega$-dependent matrix. 

Next, the 4 boundary conditions (at the left and right boundaries for the two
components, $m_x$ and $m_y$) are used to reduce the number of equations to $2N-4$.
This is performed by algebraically solving the discretized boundary conditions with
respect to the boundary points, i.e., by determining $m_i$ where $i\in\{1,N,N+1,2N\}$
in terms of the magnetizations at the interior points. 

Finally, each $(2N-4)\times (2N-4)$ matrix is separated into two parts: a term
independent of the frequency $\omega$ and a term proportional to $\omega$. The dipole
interaction causes the eigenvalue problem to be non-Hermitian and therefore
computationally more demanding than a generalized eigenvalue problem. We find the
dispersion relation and magnetization vectors by solving this eigenvalue problem. The
resulting eigenvectors are used to find the magnetization at the boundary by
back-substitution into the equations for the boundary conditions.

We are interested in finding the mode and wave-vector dependence of the spin-pumping
enhanced Gilbert damping. To obtain this information numerically, we perform two
independent calculations of the (complex) eigenvalues. First, we calculate the
complex eigenvalues $\omega_\text{d}$ when there is no spin pumping, but dissipation
occurs via the conventional bulk Gilbert damping. Second, we calculate the complex
eigenvalues $\omega_\text{sp}$ when spin pumping is active at the FI-NM interface 
but there is no bulk Gilbert damping. A mode- and wave-vector-dependent measure of
the effective enhanced Gilbert damping enhancement is then given by 
\begin{equation}
    \Delta\alpha=\alpha\frac{\Im\omega_\text{sp}}{\Im\omega_\text{d}}.
\end{equation}
To ensure that we treat the same modes in the two independent calculations, we check
the convergence of the relative difference in the real part of the eigenvalues.
Table~\ref{tab:parameters} lists the values for the different system parameters that
are used throughout this section.

Let us first discuss the renormalization of the Gilbert damping when there is no
surface anisotropy. We will present the numerical results for the three main
geometries described in Sec.~\ref{sec:introduction} and compare the results to the
analytical results of Sec.~\ref{sec:long_wavelength_modes}.

\subsection{FVMSW (\texorpdfstring{$\theta=0$}{theta=0})}
%--------------------------------------------------------------------
\begin{figure}[htbp]
    \includegraphics[width=\linewidth]{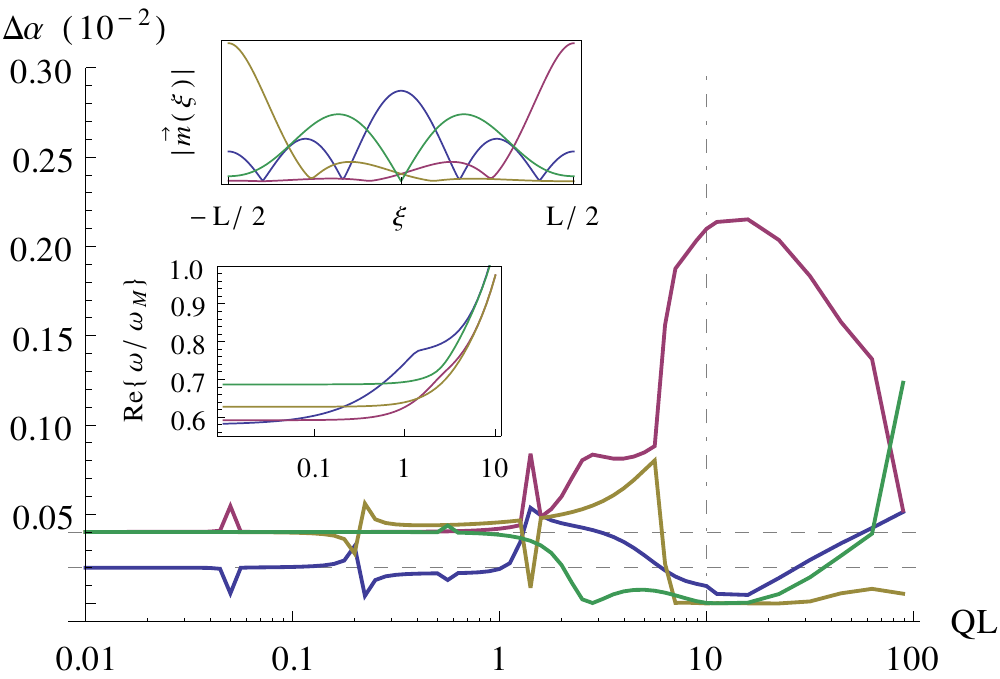}
    \caption{\label{fig:deltaAlphaFVMSW}$\Delta\alpha$ versus wave vector for the
    FVMSW geometry of the four smallest eigenvalues. Top inset: Magnitudes
    of eigenvectors (in arbitrary units) across the film at $QL=10$. Bottom inset:
    dispersion relation in the dipole-dipole active regime.}
\end{figure}
%--------------------------------------------------------------------
Figure~\ref{fig:deltaAlphaFVMSW} shows the wave-vector dependent renormalization of
the Gilbert damping $\Delta \alpha$ due to spin pumping at the FI-NM interface in the
FVMSW geometry. In this geometry, waves travelling along $\pm\buvec\zeta$ have the
same symmetry; thus, each line is doubly degenerate and corresponds to two waves of
$\pm\omega$. The ``spikes'' in the figure are due to degeneracies, i.e., mode
crossings, and upon inspection, these spikes can be observed in the dispersion
relation.
%--------------------------------------------------------------------

\subsubsection{Easy-Axis Surface Anisotropy (\texorpdfstring{$\buvec\xi$}{xi} easy 
axis)}
%--------------------------------------------------------------------
\begin{figure}[thbp]
    \includegraphics[width=\linewidth]{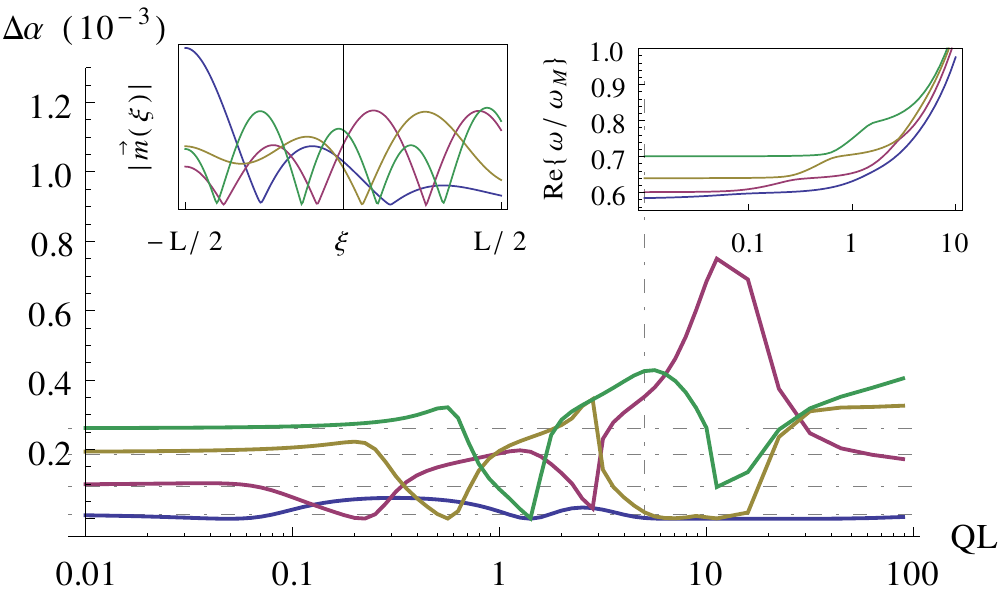}
    \caption{\label{fig:deltaAlphaFVMSW_EASA}$\Delta\alpha_\text{EA}$ versus wave
    vector for the FVMSW geometry showing the four smallest eigenvalues. The horizontal
    dashed lines indicate solutions of Eq.~\eqref{eq:deltaAlpha_EA_perp}. Left
    inset: Magnitudes of eigenvectors (in arbitrary units) across the film at $QL=5$.
    Right inset: Dispersion relation in the dipole-dipole active regime.}
\end{figure}
%--------------------------------------------------------------------
Figure~\ref{fig:deltaAlphaFVMSW_EASA} shows $\Delta\alpha_\text{EA}$ for the FVMSW geometry
with an EA surface anisotropy at the spin-active interface. As predicted in
Sec.~\ref{sec:EA_perp}, all modes exhibit a decreased $\Delta\alpha$ compared with
those in Eqs.~\eqref{eq:deltaAlpha_0} and \eqref{eq:deltaAlpha_n}. For small $QL$ and
the chosen value of $K_s$ (see Tab.~\ref{tab:parameters}), the $1^\text{st}$ four
modes match the analytical result of Eq.~\eqref{eq:deltaAlpha_EA_perp}, which is
consistent with the plot in Figure \ref{fig:EA_dissipation_factor}. For even higher
excited modes, the effect of the EA surface anisotropy becomes weaker due to the
increase in transverse exchange energy. These modes (not shown in the figure)
approach the value of $\Delta\alpha_n$.

\subsubsection{Easy-Plane Surface Anisotropy (\texorpdfstring{$\buvec\xi$}{xi} hard
axis)\label{sssec:FVMSW_EPSA}}
%--------------------------------------------------------------------
\begin{figure}[h!tb]
    \includegraphics[width=\linewidth]{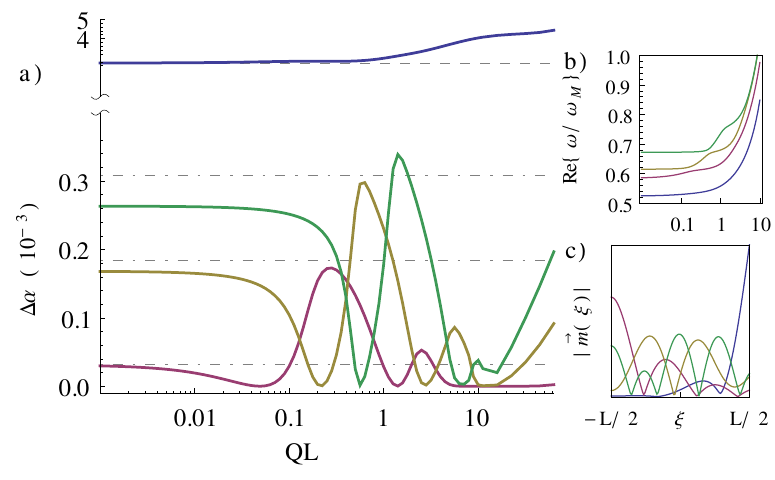}
    \caption{\label{fig:deltaAlphaFVMSW_EPSA}a) $\Delta\alpha_\text{EP}$ versus wave
    vector for the FVMSW geometry, showing the four smallest eigenvalues. The dashed
    lines represent the analytic solutions from Sec.~\ref{sec:oop_EP}. b) Dispersion
    relation in the dipole-dipole active regime. c) Magnitude of eigenvectors (in
    arbitrary units) across the film at $QL=5$.}
\end{figure}
%--------------------------------------------------------------------
Figure~\ref{fig:deltaAlphaFVMSW_EPSA} shows $\Delta\alpha_\text{EP}$ for the FVMSW
geometry with an EP surface anisotropy. We see that the mode corresponding to $n=0$
has been promoted to a surface mode with a large $\Delta\alpha$, which for small
values of $QL$ matches Eq.~\eqref{eq:deltaAlpha_FVMSW_EP_0}. For the higher excited
modes, we observe a decrease in $\Delta\alpha$ compared to the case with no surface
anisotropy. 

\subsection{BVMSW (\texorpdfstring{$\theta=\pi/2$}{theta=pi/2} and
\texorpdfstring{$\phi=0$}{phi=0})}
%--------------------------------------------------------------------
\begin{figure}[h!bt]
    \includegraphics[width=\linewidth]{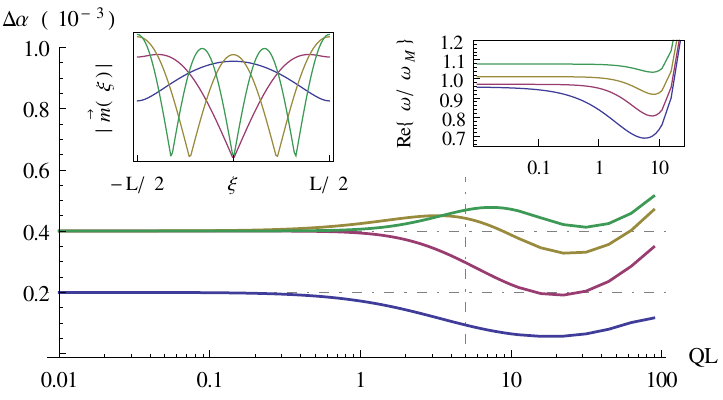}
    \caption{\label{fig:deltaAlphaBVMSW}$\Delta\alpha$ versus wave vector for the
    BVMSW geometry ($\theta=\pi/2$ and $\phi=0$) with $K_s=0$, plotted for the four
    smallest eigenvalues. Left inset: magnitudes of normalized eigenvectors across the
    film at $QL=5$. Right inset: dispersion relation in the dipole-dipole active
    regime.}
\end{figure}
%--------------------------------------------------------------------
Figure~\ref{fig:deltaAlphaBVMSW} shows the $QL$-dependent renormalization of the
Gilbert damping due to spin pumping at the FI-NM interface in the BVMSW geometry. We
see that the enhancement $\Delta\alpha$ agrees with the analytic limits in
Eqs.~\eqref{eq:deltaAlpha_0} and \eqref{eq:deltaAlpha_n} for small values of $QL$.
For large values of $QL$, we are in the strong exchange regime, in which the in-plane
exchange energy becomes large compared to all other energy contributions. This
in-plane exchange stiffness effectively quenches the coupling to the normal metal
layer, causing $\Delta\alpha\to0$ for large values of $QL$.

Although Figure~\ref{fig:deltaAlphaBVMSW} only appears to show the three first
eigenvalues and eigenvectors, it actually contains double this amount. Because
$\buvec z$ is parallel to the wave-propagation direction $\buvec\zeta$ in this
geometry, there is no change in dipolar energies, regardless of whether the wave
travels in the $+\buvec\zeta$ direction or in the $-\buvec\zeta$ direction; thus, the
Gilbert damping is enhanced equally in both wave directions. A slight offset from
this configuration, taking either $\theta<\pi/2$ or $\phi\neq0$, would result in a
splitting of each line in Figure~\ref{fig:deltaAlphaBVMSW} into two distinct lines.

\subsubsection*{Including Surface Anisotropy}
%--------------------------------------------------------------------
\begin{figure*}[hbtp]
    \centering
    \begin{tabular}{ll}
        \includegraphics[width=0.5\linewidth]{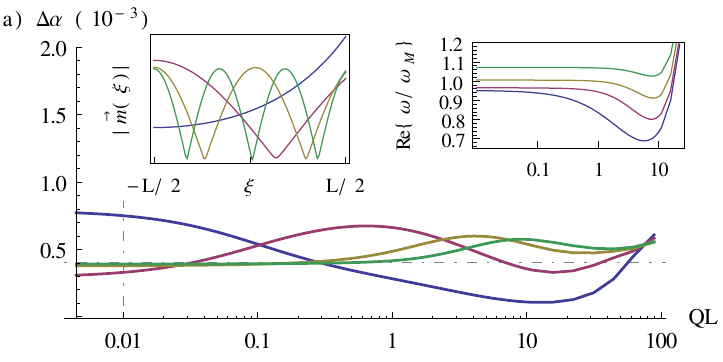}
        &
        \includegraphics[width=0.5\linewidth]{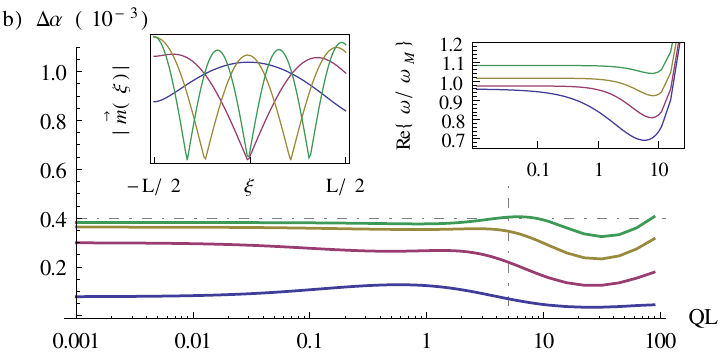}
    \end{tabular}
    \caption{\label{fig:deltaAlphaBVMSW_EAEP}a) Dispersion relation versus wave vector for
    the BVMSW geometry ($\theta=\pi/2$, $\phi=0$) for the four lowest eigenvalues in
    the case of EA surface anisotropy. b) Dispersion relation in the case of EP
    surface anisotropy. In both figures, the horizontal dashed lines mark the value
    of $\Delta\alpha_n$ in the case of no surface anisotropy.}
\end{figure*}
%--------------------------------------------------------------------
Figure~\ref{fig:deltaAlphaBVMSW_EAEP} shows both the EA and the EP surface-anisotropy
calculations in the BVMSW geometry. In the case of an EA surface anisotropy, the mode
corresponding to $n=0$ gets promoted to a surface mode, similarly to the case in
which there is EP surface anisotropy in the FVMSW geometry. The increase in
$\Delta\alpha$ is much smaller for the same magnitude of $K_s$, as explained in
detail in Sec.~\ref{sec:SA_incl}. The higher modes, corresponding to $n>0$, exhibit
increased quenching of the Gilbert damping enhancement. In the case of EP surface
anisotropy, all modes exhibit quenched Gilbert damping enhancement.

\subsection{MSSW (\texorpdfstring{$\theta=\phi=\pi/2$}{theta=phi=pi/2})}
%--------------------------------------------------------------------
\newsavebox{\tempbox}
\newsavebox{\tempboxtwo}
\begin{figure*}[hbtp]
    \centering
    \sbox{\tempbox}{
        \includegraphics[width=0.5\linewidth]{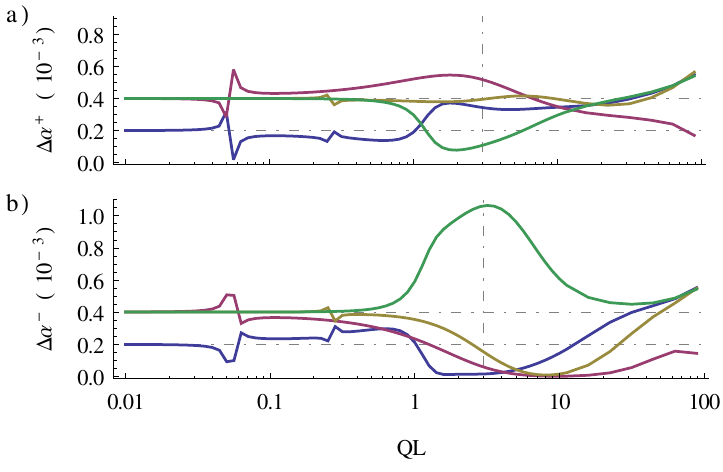}
    }
    \sbox{\tempboxtwo}{
    \vbox to\ht\tempbox{\hsize=0.5\linewidth
        \vfil
        \includegraphics[width=0.5\linewidth]{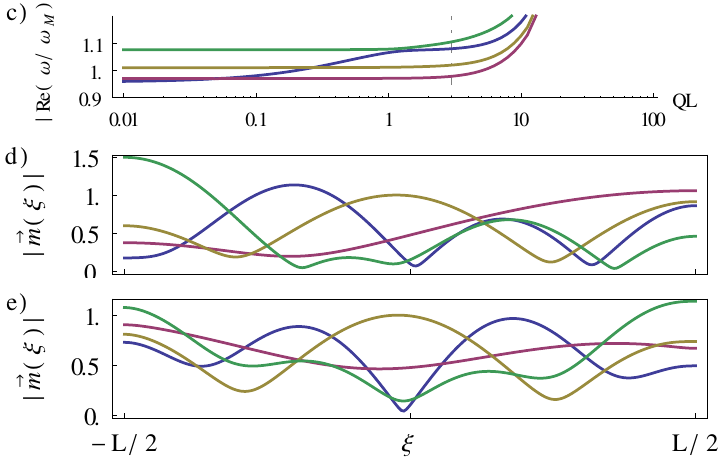}
        \vfil
        }}
    \begin{tabular}{ll}
        \usebox{\tempbox} & \usebox{\tempboxtwo}
    \end{tabular}
    \caption{\label{fig:deltaAlphaMSSW}Gilbert damping renormalization in the MSSW
    geometry. Subplots a) and b) show Gilbert damping renormalization $\Delta\alpha$
    for modes with positive (negative) $\Re\{\omega\}$. The horizontal dashed lines
    represent the analytical values $\Delta\alpha_0$ and $\Delta\alpha_n$ for small
    $QL$. c) Dispersion relation versus wave vector for the MSSW geometry
    ($\theta=\phi=\pi/2$) for the four smallest eigenvalues, colored pairwise in
    $\pm\omega$. Subplot d (e) shows the magnitude of normalized eigenvectors (in
    arbitrary units) at $QL=3$ across the film modes with positive (negative)
    $\Re\{\omega\}$.}
\end{figure*}
%--------------------------------------------------------------------
Figure~\ref{fig:deltaAlphaMSSW} shows the $QL$-dependent renormalization of the
Gilbert damping due to spin pumping at the FI-NM interface in the MSSW geometry. The
computed eigenvalues agree with Eqs.~\eqref{eq:deltaAlpha_0} and
\eqref{eq:deltaAlpha_n} for small values of $QL$. We see in the inset of
Figure~\ref{fig:deltaAlphaMSSW} that in this geometry, the macrospin-like mode
behaves as predicted by \citet{Damon1961308}\cite{PhysRev.118.1208}, cutting through
the dispersion relations of the higher excited modes for increasing values of $QL$ in
the dipole-dipole regime. A prominent feature of this geometry is the manner in which
the modes with different signs of $\Re\{\omega\}$ behave differently due to the
dipole-dipole interaction. This is because the internal field direction ($\buvec{z}$)
is not parallel to the direction of travel ($\buvec\zeta$) of the spin wave. Hence,
changing the sign of $\omega$ is equivalent to inverting the externally applied
field, changing the $xyz$ coordinate system in Figure~\ref{fig:geometry} from a
right-handed coordinate system to a left-handed system. In the middle of the dipole
regime, the lack of symmetry with respect to propagation direction has different
effects on the eigenvectors; e.g., in the dipole-dipole active region the modes with
positive or negative $\Re\{\omega\}$ experience an increased or decreased magnitude
of the dynamic magnetization, depending on the value of $QL$, as shown in
Figure~\ref{fig:deltaAlphaMSSW}e \& f. This magnitude difference creates different
renormalizations of the Gilbert damping, as the plot of $\Delta\alpha^{(\pm)}$ in
Figure~\ref{fig:deltaAlphaMSSW}b \& c shows.

\subsubsection*{Including Surface Anisotropy}
%--------------------------------------------------------------------
\begin{figure*}[h!tb]
    \centering
    \begin{tabular}{rr}
        \includegraphics[width=0.5\linewidth]{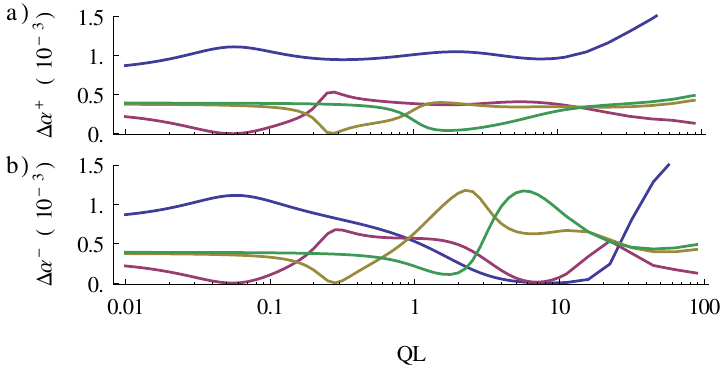}
        &
        \includegraphics[width=0.5\linewidth]{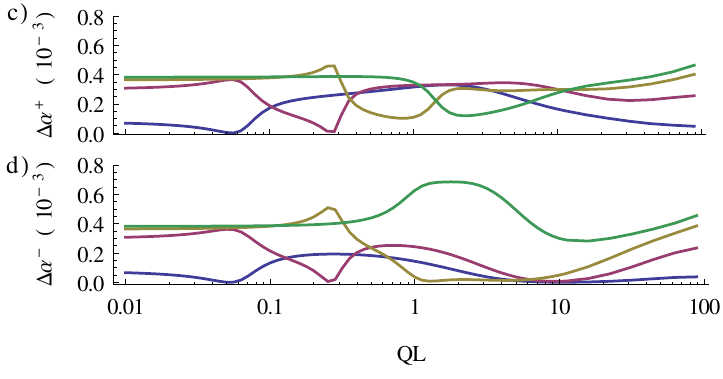}
    \end{tabular}
    \caption{\label{fig:deltaAlphaMSSW_EA}a) and b) Gilbert damping renormalization
    from spin pumping in the MSSW geometry ($\theta=\phi=\pi/2$) for modes with
    positive (negative) $\Re\{\omega\}$ in the case of EA surface anisotropy. The
    four smallest eigenvalues are colored pairwise in $\pm\omega$ across the plots.
    c) and d) show the Gilbert damping renormalization in the case of EP surface
    anisotropy.}
\end{figure*}
%--------------------------------------------------------------------
Figure~\ref{fig:deltaAlphaMSSW_EA} shows $\Delta\alpha$ computed for modes in the
MSSW geometry with EA and EP surface anisotropies. We can clearly see that for small
$QL$ an exponentially localized mode exists in the EA case, and as predicted in
Sec.~\ref{sec:SA_incl}, all the lowest-energy modes have spin pumping quenched by EP
surface anisotropy. This is similar to the corresponding case in the BVMSW geometry.

\subsection{AC and DC ISHE}
%--------------------------------------------------------------------
\begin{figure}
    \centering
    \includegraphics[width=\linewidth]{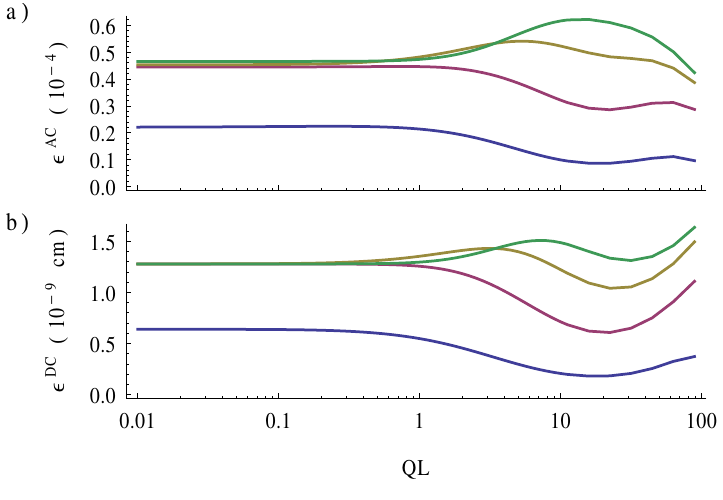}
    \caption{\label{fig:BVMSW_ISHE_MEASURES}ISHE as a function of in-plane wave vector in the BVMSW geometry with
    $K_s=0$. a) AC ISHE measure of Eq.~\eqref{eq:Eishe_dc_normalized}; b) DC ISHE
    measure of Eq.~\eqref{eq:Eishe_dc_normalized}.}
\end{figure}
%--------------------------------------------------------------------
Figure \ref{fig:BVMSW_ISHE_MEASURES} shows the DC and AC ISHE measures for the BVMSW
geometry corresponding to the data represented in Figure \ref{fig:deltaAlphaBVMSW}.
In this geometry, the angular term, $\sin\theta$, in
Eq.~\eqref{eq:Eishe_dc_normalized} is to equal one, ensuring that the DC measure is
nonzero. This is not the case for all geometries because the DC electric field
vanishes in the FVMSW geometry. The mode-dependent DC ISHE measure exhibits the same
$QL$-dependence as the spectrum of the Gilbert damping enhancement in all geometries
where $\sin\theta\neq0$. We have already presented the renormalization of the Gilbert
damping in the most general cases above. Therefore, we restrict ourselves to
presenting the simple case of the BVMSW geometry with no surface anisotropy here.

The AC ISHE measure plotted in Figure~\ref{fig:BVMSW_ISHE_MEASURES} exhibits a
similar $QL$ dependence to the Gilbert damping renormalization (and hence the DC ISHE
measure), but with a slight variation in the spectrum towards higher values of $QL$.
Note that because Eq.~\eqref{eq:Eishe_ac} is non-zero for all values of $\theta$, the
AC effect should be detectable in the FVMSW geometry. By comparing the computed
renormalization of the Gilbert damping for the different geometries in the previous
subsections, we see that the strong renormalization of the $n=0$ induced surface mode
that occurs in the FVMSW geometry with easy-plane surface anisotropy (see
Sec.~\ref{sssec:FVMSW_EPSA} and Fig.~\ref{fig:deltaAlphaFVMSW_EPSA}) can have a
proportionally strong AC ISHE signal in the normal metal.

\section{Conclusion\label{sec:conclusion}}
In conclusion, we have presented analytical and numerical results for the
spin-pumping-induced Gilbert damping and direct- and alternating terms of the inverse
spin-Hall effect. In addition to the measures of the magnitudes of the DC and AC
ISHE, the effective Gilbert damping constants strongly depend on the modes through the
wave numbers of the excited eigenvectors.

In the long-wavelength limit with no substantial surface anisotropy, the spectrum is
comprised of standing-wave volume modes and a uniform-like (macrospin) mode. These
results are consistent with our previous findings\cite{PhysRevLett.111.097602}: in
the long-wavelength limit, the ratio between the enhanced Gilbert damping for the
higher volume modes and that of the macrospin mode is equal to two. When there is
significant surface anisotropy, the uniform mode can be altered to become a pure
localized surface mode (in the out-of-plane geometry and with EP surface
anisotropy), a blend between a uniform mode and a localized mode (in-plane geometries
and EA surface anisotropy), or quenched uniform modes (out-of-plane field
configuration and EA surface anisotropy, or in-plane field configuration and EP
surface anisotropy). The effective Gilbert damping is strongly enhanced for the
surface modes but decreases with increasing surface-anisotropy energies for all the
other modes.

The presented measures for both the AC and DC inverse spin-Hall effects are strongly
correlated with the spin-pumping renormalization of the Gilbert damping, with the DC
effect exhibiting the same $QL$ dependency, whereas the AC effect exhibits a slighthly
different variation for higher values of $QL$. Because the AC effect is nonzero in
both in-plane and out-of-plane geometries and because both EP and EA surface
anisotropies induce surface-localized waves at the spin-active interface, the AC ISHE
can be potentially large for these modes.

\begin{acknowledgments}
    We acknowledge support from EU-FET grant no. 612759 (``InSpin''), ERC AdG grant
    no. 669442 (``Insulatronics''), and the Research Council of Norway grant no.
    239926.
\end{acknowledgments}

\appendix
\section{Coordinate transforms\label{app:coord-trans}}
The transformation for vectors from $\xi\eta\zeta$ to $xyz$ coordinates (see
Fig.~\ref{fig:geometry}) is given by an affine transformation matrix $T$, so that
\[
    \bvec f_{(xyz)}=\bvec T\cdot\bvec f_{(\xi\eta\zeta)},
\]
for some arbitrary vector $\bvec f$. Tensor--vector products are transformed by
inserting a unity tensor $\bvec I=\bvec T^{-1}\bvec T$ between the tensor and
vector and by left multiplication by the tensor $\bvec T$, such that the tensor
transforms
as $\bvec T\widehat{\mathcal{G}}\bvec T^{-1}$ for some tensor $\widehat{\mathcal{G}}$
written in the $\xi\eta\zeta$ basis.

$\bvec T$ is given by the concatenated rotation matrices $\bvec T=\bvec
R_{2}\cdot\bvec R_{1}$, where $R_{1}$ is a rotation $\phi$ around the $\xi$-axis, and
$R_{2}$ is a rotation $\theta-\frac{\pi}{2}$ around the new $\eta$-axis/$y$-axis.
Hence,
\begin{gather}
    \bvec R_{1}=\begin{pmatrix}
        1 & 0 & 0\\
        0 & \cos\phi & -\sin\phi\\
        0 & \sin\phi & \cos\phi
    \end{pmatrix},\\
    \begin{split}
        \bvec R_{2}=&\begin{pmatrix}
        \sin\theta & 0 & -\cos\theta\\
        0 & 1 & 0\\
        \cos\theta & 0 & \sin\theta
    \end{pmatrix},\end{split}
\end{gather}
such that
\begin{align}
    \bvec T= & \begin{pmatrix}
        \sin\theta & -\cos\theta\sin\phi & -\cos\theta\cos\phi\\
        0 & \cos\phi & -\sin\phi\\
        \cos\theta & \sin\theta\sin\phi & \sin\theta\cos\phi
    \end{pmatrix}.\label{eq:coordinate_transformation}
\end{align}
This transformation matrix consists of orthogonal transformations;
thus, the inverse transformation, which transforms $xyz\to\xi\eta\zeta$,
is just the transpose, $\bvec T^{-1}=\bvec T^{T}$.
%--------------------------------------------------------------------
%\bibliography{library}

\end{document}